\documentclass[a4paper,nofootinbib,superscriptaddress,
secnumarabic,showpacs,eqsecnum]{revtex4}

\usepackage[T1]{fontenc}

\usepackage{amsmath, amsthm, amsfonts, amssymb, latexsym}
\usepackage{graphicx}
\usepackage{color}
\usepackage{epsfig}

\usepackage[a4paper]{geometry}

\usepackage[caption=false]{subfig}

\usepackage{url}
\usepackage{ifpdf}

\ifpdf
\usepackage[unicode,pdftex]{hyperref}
\hypersetup{
    unicode=true,          
    pdftitle={Thermodynamical description of stationary, asymptotically flat solutions with conical singularities},
    pdfauthor={Carlos Herdeiro},
    pdfsubject={General Relativity},
    pdfcreator={LaTeX},          
    pdfproducer={LaTeX},         
    pdfkeywords={General Relativity, Exact solutions,Black Holes},
    colorlinks=false,            
  }
\else
  \usepackage[dvips]{hyperref}
\fi

\def\bequ{\begin{equation}}
\def\eequ{\end{equation}}

\begin{document}

\title{\Large Thermodynamical description of stationary, asymptotically flat solutions with conical singularities}

\author{Carlos Herdeiro}\email{crherdei@fc.up.pt}
\affiliation{
  Centro de F\'\i sica do Porto -- CFP \\ 
  Departamento de F\'\i sica e Astronomia \\ 
  Faculdade de Ci\^encias da Universidade do Porto -- FCUP \\
  Rua do Campo Alegre, 4169-007 Porto, Portugal
}

\author{Eugen Radu}\email{radu@theorie.physik.uni-oldenburg.de}
\affiliation{
  Institut f\"ur Physik, Universit\"at Oldenburg \\
   Postfach 2503 D-26111 Oldenburg, Germany
}

\author{Carmen Rebelo}\email{mrebelo@fc.up.pt}
\affiliation{
  Centro de F\'\i sica do Porto -- CFP  \\
  Departamento de F\'\i sica e Astronomia \\
  Faculdade de Ci\^encias da Universidade do Porto -- FCUP \\
  Rua do Campo Alegre, 4169-007 Porto, Portugal
}

\date{April 2010} 

\begin{abstract}
We examine the thermodynamical properties of a number of asymptotically flat, 
stationary (but not static) solutions having conical singularities, with both connected and non-connected event horizons, using the  thermodynamical description recently proposed in \cite{Herdeiro:2009vd}. The examples considered are the double-Kerr solution, the black ring rotating in either $S^2$ or $S^1$ and the black Saturn, where the balance condition is not imposed for the latter two solutions. We show that not only the Bekenstein-Hawking area law is recovered from the thermodynamical description but also the thermodynamical angular momentum is the ADM angular momentum.  We also analyse the thermodynamical stability and show that, for all these solutions, either the isothermal moment of inertia or the specific heat at constant angular momentum is negative, at any point in parameter space. Therefore, all these solutions are thermodynamically unstable in the grand canonical ensemble.
\end{abstract}


\maketitle

\section{Introduction}
The existence of conical singularities on a manifold implies geodesic incompleteness and therefore these singularities must be regarded as space-time boundaries from the viewpoint of classical differential geometry. Semi-classically, however, solutions with conical singularities have generically a well defined Euclidean action, and thus well defined thermodynamical properties \cite{Gibbons:1979nf}. This suggests that such solutions may be associated to well defined quantum states in an appropriate theory of quantum gravity, upon a correct 
quantisation of the physical parameters, including the conical deficit/excess. 
Indeed there are special examples of manifolds with conical singularities in which a quantum theory of gravity is well defined. This is the case of string theory on certain orbifold \cite{Dixon:1985jw,Dixon:1986jc}; in these backgrounds string theory is solvable and smooth, due to new degrees of freedom localised at the tip of the cone - the twisted sector. 

Even at the classical level, there is usually a clear physical interpretation for the conical singularity. For instance, as providing an otherwise impossible mechanical balance between two black holes. In this class of examples, studying the singularity's properties unveils features about black hole interactions, as shown in  \cite{Herdeiro:2008kq,Costa:2009wj} in the case of the double Kerr solution \cite{dietz}
 in four spacetime dimensions. 
Thus, despite their incomplete nature, backgrounds with conical singularities
have well defined physical properties worth investigating.

In this paper we shall continue to investigate the thermodynamical description introduced in  \cite{Herdeiro:2009vd}. 
The basic observation therein was that a natural choice of thermodynamical variables leads to the Bekenstein-Hawking area law for the entropy, even for solutions with non-connected event horizons, in contrast to previous approaches. All the examples given in \cite{Herdeiro:2009vd} were static solutions. Here we consider various stationary solutions which,  similarly to \cite{Herdeiro:2009vd},  are asymptotically flat, i.e the conical singularity does not extend to spatial infinity.
 We show that not only the area law is obtained for the entropy, but also that the 
 angular momentum computed thermodynamically from the free energy coincides with the 
 ADM angular momentum, unlike the thermodynamical description in  \cite{Costa:2009wj}. 
 Since we have a consistent thermodynamical description of these solutions, 
 we can also examine their thermodynamical stability. 
 The generic feature found is that they are all unstable in the grand canonical ensemble. 

This paper is organised as follows. In section 2 we introduce the appropriate thermodynamical variables, the various energy functionals and all relevant formulas to be applied to the specific examples. These are considered in the following sections: black rings in section 3, the double Kerr in section 4 and the black Saturn in section 5. We present some final remarks in section 6.

\section{Formalism}
\label{formalism}
In order to discuss the thermodynamical properties of such backgrounds an appropriate set of thermodynamical variables must be chosen. It was suggested in \cite{Herdeiro:2009vd} that the first law of thermodynamics for vacuum, stationary, asymptotically flat solutions with conical singularities reads
\begin{equation}
    d\mathcal{M}=T_{H}dS + \Omega dJ + P d\mathcal{A} \ , \qquad \mathcal{M}=\mathcal{M}[S,J,\mathcal{A}] \ . \label{fl1}
\end{equation} 
The first two terms on the right hand side are the standard ones also present for regular solutions. These involve the Hawking temperature, $T_H$, the entropy, $S$, the angular velocity of the horizon (or horizons), $\Omega$ and the angular momentum, $J$. In the presence of a conical singularity, which is exerting a pressure $P$, with world-volume spanning a spacetime area $\mathcal{A}/T_H$ (computed in the Euclidean section), the last term should be included. Moreover, the energy functional $\mathcal{M}$ is the ADM mass, $M_{ADM}$, subtracted by the energy associated with the conical singularity as seen by an asymptotic static observer, $E_{int}$. Generically, $E_{int}=-P\mathcal{A}$ \cite{Herdeiro:2009vd}. Transforming to the energy functional $M_{ADM}$, corresponds therefore to a Legendre transform 
\bequ
M_{ADM}=\mathcal{M}-P\mathcal{A} \ , \label{thermomass}\eequ
leading to the first law in the form:
\begin{equation}
    dM_{ADM}=T_{H}dS + \Omega dJ - \mathcal{A}dP \ , \qquad M_{ADM}=M_{ADM}[S,J,P] \ . \label{fl2}
\end{equation} 

Observe that for the energy functional $\mathcal{M}$ the independent variables are the extensive ones. That $\mathcal{A}$, rather than $P$, is the extensive variable can be seen from an example. Consider the $\mathbb{Z}_2$ invariant Israel-Khan solution with three Schwarzschild black holes. The conical singularity (and hence $P$) is the same in the two conical sub-sets, and the total $\mathcal{A}$ is twice that of each conical sub-set. It follows that thermodynamical equilibrium requires that $T_H,\Omega$ are the same in all connected components of the event horizon and that $P$ is the same in all sub-spaces where there exist conical singularities. Of course, if there are no conical singularities, the two `internal energy' functionals are equal, $M_{ADM}=\mathcal{M}$, and \eqref{fl1} and \eqref{fl2} coincide. But generically, $\mathcal{M}$ plays a more fundamental role. This seems natural; indeed $\mathcal{M}$, unlike $\mathcal{M}_{ADM}$, takes into account the Komar masses of the individual black objects \textit{and} the energy associated to the conical singularity. Thus, the remaining energy functionals (grand canonical potential, canonical potential and enthalpy) are obtained by the standard Legendre transforms from the energy functional $\mathcal{M}$. 

Alternatively, if the Legendre transform starts from the ADM mass, it must also include the variables associated to the conical singularity. For instance, performing the Legendre transform, 
\begin{equation}
W=\mathcal{M}-T_HS-\Omega J=M_{ADM}-T_HS-\Omega J+P\mathcal{A} \ , \label{w} \end{equation}
we obtain the first law in terms of the grand canonical potential $W$ (or Gibbs free energy)
\bequ
dW=-S dT_{H}  - J d\Omega  + P d\mathcal{A} \ , \qquad W=W[T_H,\Omega, \mathcal{A}] \ .\label{firstgibbs}\eequ
Thus, the entropy $S$, angular momentum $J$, pressure $P$, and energy functional $\mathcal{M}$ of the system are given by
\begin{equation}
    S=-\frac{\partial{W}}{\partial{T_H}}\Bigg|_{\Omega,\mathcal{A}}\ , \qquad J=-\frac{\partial{W}}{\partial{\Omega}}\Bigg|_{T_{H},\mathcal{A}}\ , \qquad 
P=\frac{\partial{W}}{\partial{\mathcal{A}}}\Bigg|_{T_{H},\Omega} \ . \label{thermorelation}
\end{equation}
In the Euclidean approach to black hole thermodynamics, the grand canonical potential is determined by 
\bequ
W=T_H I \ , \label{gc}
\eequ
where $I$ is the Euclidean action. This action is
\bequ 
I=I_0-\frac{E_{int}}{T_H} \ .
 \label{action} 
 \eequ
$I_0$ is the contribution to the action from black objects found when neglecting the conical singularity (which arises from the boundary term). Consider the Euclidean action  for the metric $g$ over a region $Y$ with a boundary $\partial Y$ to have the form  \cite{Gibbons:1976ue}:
\begin{equation}
    I=-\frac{1}{16\pi G_d}\int_Y \,R \, - \frac{1}{8\pi G_d}\oint_{\partial Y} \,(K-K_0)\ , \label{incompleteaction}
\end{equation}
where $K$ ($K_0$) is the trace of the second fundamental form of the boundary $\partial Y$ 
(embedded in flat space) and $G_d$ is
Newton's constant in $d-$dimensions.\footnote{In the following we take units in which $G_d=1$.} An asymptotically flat Euclidean metric with a single angular momentum parameter may be written in the form, near spatial infinity,\footnote{A higher dimensional black object
may have $k=[(d-1)/2]$ different angular momenta. Known examples with $k>1$ are  the
$d>4$ Myers-Perry black hole \cite{Myers:1986un} and the $d=5$ Pomeransky-Sen'kov double
spinning black ring \cite{Pomeransky:2006bd}. The generalisation of our results to higher $k$ is  straightforward.}
\begin{eqnarray}
 ds^2\thickapprox \left(1-\frac{\mu}{r^{d-3}} \right)d\tau ^2+\left(1+\frac{\mu}{r^{d-3}} \right)
  dr^2+r^2(d\theta^2 +\sin^2\theta d\phi^2+\cos^2\theta d\Omega^2_{d-4})  
  + \frac{2j\sin^2 \theta}{r^{d-3}} d\tau d\phi ,~~~{~~~~}
\end{eqnarray}
where the parameters $\mu,j$ relate to the ADM mass and angular momentum
as
\begin{eqnarray}
\mu = \frac{16 \pi\,M_{ADM}}{(d-2)\Omega_{d-2} } \ ,~~
|j|=\frac{8 \pi\,J_{ADM}}{ \Omega_{d-2} } \ ;
\end{eqnarray}
$\Omega_{d-2}$ denotes the area of the unit $(d - 2)$-sphere.

The extrinsic curvature is defined as $K_{\mu \nu}=-h^\alpha_\mu h^\beta_\nu \nabla_\alpha n^\beta,$ where  $h_{\mu \nu}=g_{\mu \nu}-n_\mu n_\nu.$ Taking $\partial Y$ to be the product of the time axis with a sphere of large radius $r_0$, the unit normal to the hyper-surface $\partial Y$ is $n_\mu={\partial_\mu r}/{\sqrt{g^{rr}}}$.  The traces we need are then 
$$K=g^{\mu \nu}K_{\mu \nu}=\frac{d-2}{r_0}-\frac{\mu}{2\,r_0^{d-2}} +\mathcal{O}\left(\frac{1}{r_0^{d-1}}\right) $$ 
and $K_0=(d-2)/r_0$. Considering a space free of conical singularities; then the tree level Euclidean action will come entirely from the surface term, which gives
\begin{eqnarray*}
I_0=  \frac{M_{ADM}}{d-2} \int d\tau =\frac{M_{ADM}}{d-2} \beta \ ,
 \end{eqnarray*}
where  $\beta=1/T_H$ is the periodicity of the Euclidean time.

 The second term in \eqref{action} is given by \cite{Herdeiro:2009vd}
\bequ  E_{int}= -P\mathcal{A}= \frac{\delta}{8\pi}\mathcal{A} \ ; \label{pressure} \eequ
it is the total energy associated to the strut as seen by a static observer placed at infinity. Here $\delta$ is the conical deficit/excess associated to the conical singularity. Thus, the Gibbs free energy is always given by
\bequ
W(T_H,\Omega,\mathcal{A})=\frac{M_{ADM}(T_H,\Omega,\mathcal{A})}{d-2}-\frac{\delta(T_H,\Omega,\mathcal{A})}{8\pi}\mathcal{A} \ . \label{gibbs} \eequ
 It is now straightforward to compute the quantities \eqref{thermorelation}. These will obey the
formula
\bequ
\frac{d-3}{d-2}M_{ADM}=T_H S+\Omega J \ , \label{smarr}Ê\eequ
which is a generic consequence of  relations \eqref{thermomass} and \eqref{w}, using also \eqref{gc} and \eqref{action}. This is \textit{the Smarr formula}; but note well that this is only the case because, in our description, $S,J$ obtained from \eqref{thermorelation} \textit{coincide} with $A_H/4,J_{ADM}$, where $A_H$ is the sum of the areas of the event horizon connected components.

We shall also compute, for the examples we consider, two other quantities, which are of relevance for the analysis of thermodynamical stability. The first one is the isothermal moment of inertia
\bequ
\epsilon_{T_H,\mathcal{A}}=\frac{\partial{J}}{\partial{\Omega}}\Bigg|_{T_H,\mathcal{A}}\ , \label{e} \eequ
which is compute in the grand canonical ensemble. The second one is the specific heat at constant angular momentum, which is computed in the canonical ensemble. Thus we introduce the canonical potential (or Helmholtz free energy) via the Legendre transform
\bequ
F=W+\Omega J \ , \eequ
such that the first law takes the form
\bequ
dF=-S dT_H+\Omega dJ+Pd\mathcal{A} \ , \qquad F=F[T_H,J,\mathcal{A}] \ . \eequ
The specific heat at constant angular momentum is computed as
\bequ
C_J=T_H \frac{\partial S}{\partial T_H}\Bigg|_{J,\mathcal{A}} \ . \label{c}Ê\eequ
The reason for working with $C_J$, rather than the specific heat at constant angular velocity $C_\Omega=T_H(\partial S/\partial T_H)|_{\Omega, \mathcal{A}}$, is the following. In the grand canonical ensemble the condition for thermodynamical stability is the positivity of the Weinhold metric, which amounts to the positivity of $\epsilon_{T_H,\mathcal{A}}$ and $C_J$, rather than $C_\Omega$ (see \cite{Costa:2009wj} and references therein).

For completeness let us note that we can introduce a fourth thermodynamical potential, the enthalpy $H$, via the Legendre transform
\bequ
H=W+T_H S \ . \eequ
The first law then reads
\bequ
dH=T_H  dS -J d\Omega +Pd\mathcal{A} \ , \qquad H=H[S,\Omega, \mathcal{A}] \ . \eequ

We close this section by remarking that although the solutions of Einstein's equations 
with (naked) conical singularities should not be faced as vacuum solutions, this does not affect the
generality of the relations derived above. Indeed, as in the simplest case of a cosmic string \cite{Vilenkin:1984ib}, the conical singularity is supported by a  matter source with a precise form for its energy momentum tensor  \cite{Fursaev:1995ef}. As discussed in \cite{Herdeiro:2009vd}, however, the inclusion of the contribution of this matter source to the total action does not change the expression (\ref{action}) of the tree level action of the system. The reason is that taking into account the matter contribution, one should also  subtract the contribution of a nontrivial background  with an equivalent source, which  consists in a finite piece of a string/strut (for $d=4$) or a deficit/excess membrane (for $d=5$) in a flat spacetime geometry. To see this explicitly, consider the total tree level action
\bequ  I=-\int_Y \,\left(\frac{R}{16\pi}+\mathcal{L}_m \right)- \frac{1}{8\pi 
}\left(\oint_{\partial Y} \,K  - \oint_{\partial Y_{\delta}} 
\,K_\delta \right)\ ,\label{fullaction} \eequ
where the matter Lagrangian $\mathcal{L}_m$ has been included and the reference background, wherein the trace of the extrinsic curvature is $K_\delta$, includes an equivalent source. The Lagrangian for the matter source that supports the conical singularity is $\mathcal{L}_m =-\delta/(8 \pi)\delta_\Sigma$  \cite{Herdeiro:2009vd}, where $\delta_\Sigma$ is a Dirac delta function with support on the world-volume of the conical singularity. To compute the last term in \eqref{fullaction} write the reference background as (for concreteness take $d=4$)
$$ ds_\delta^2=d\tau^2+\rho^2 (\delta_{-\infty}^{-\bar{z}}+a^2 
\delta_{-\bar{z}}^{\,\bar{z}}+\delta_{\bar{z}}^{\infty})d\phi^2+d\rho^2+dz^2 \, \ ,$$
where $\delta_a^b=1$  for $z\in [a,b]$ and zero otherwise.  Then, considering $\partial Y$ to be the product of the times axis with a cylinder around the $z$ axis,
\begin{eqnarray*}
     \oint_{\partial Y_{\delta}} \,K_\delta-\oint_{\partial Y} 
\,K_0 =2\pi \beta \, 
\int_{-\bar{z}}^{\,\bar{z}}\left(a-1\right)dz = - \delta Area,
\end{eqnarray*}
where $\delta=2\pi(1-a)$ is the conical deficit/excess present and $Area = 
\int_{-\bar{z}}^{\,\bar{z}} d\tau dz$ is the area of the surface
spanned by the conical singularity. Thus
 \bequ-\int_Y \,\mathcal{L}_m +\frac{1}{8\pi}\oint_{\partial 
Y_{\delta}} \,K_\delta =\frac{1}{8\pi}\oint_{\partial Y} 
\,K_0\ , \eequ which brings us back to the Euclidean action \eqref{incompleteaction}. This argument reveals, again, the physical significance of the pressure $P=-\delta/8\pi$ and of its conjugate variable ${\cal A}=Area/\beta$; they define the physical properties of the matter source supporting the conical geometry.

In the following sections we shall compute \eqref{thermorelation}, using \eqref{gibbs}, as well as \eqref{e} and \eqref{c}, for various examples.

\section{Black rings rotating in a single plane}
The rotating black ring solution in $d=5$ Einstein gravity provides perhaps the simplest application of this formalism. The thermodynamics of the static solution was investigated in \cite{Herdeiro:2009vd}
where it was argued that, in the absence of rotation, all configurations are thermodynamically
unstable. We shall now see that this conclusion still holds when rotation in either $S^1$ or $S^2$ is included.

\subsection{$S^1$ rotating black ring} 

\begin{figure}
\begin{picture}(0,0)(0,0)
\end{picture}
\centering
\includegraphics[width=0.45\textwidth]{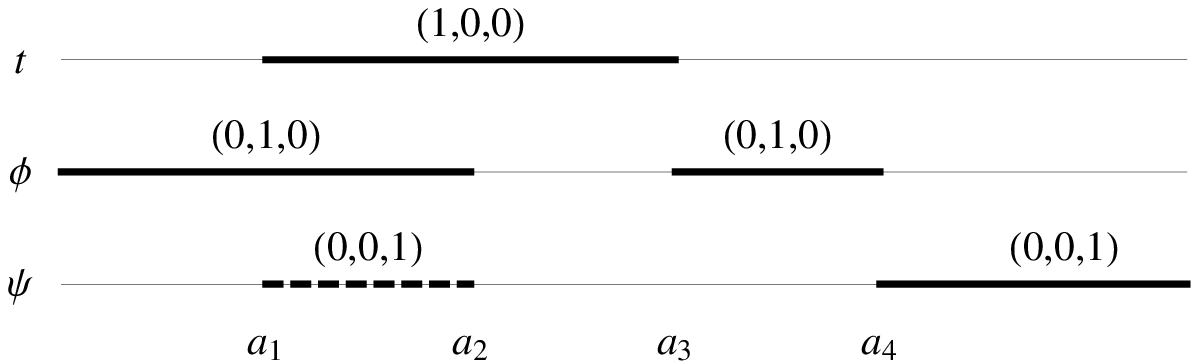} \ \ \ \ \ 
\includegraphics[width=0.45\textwidth]{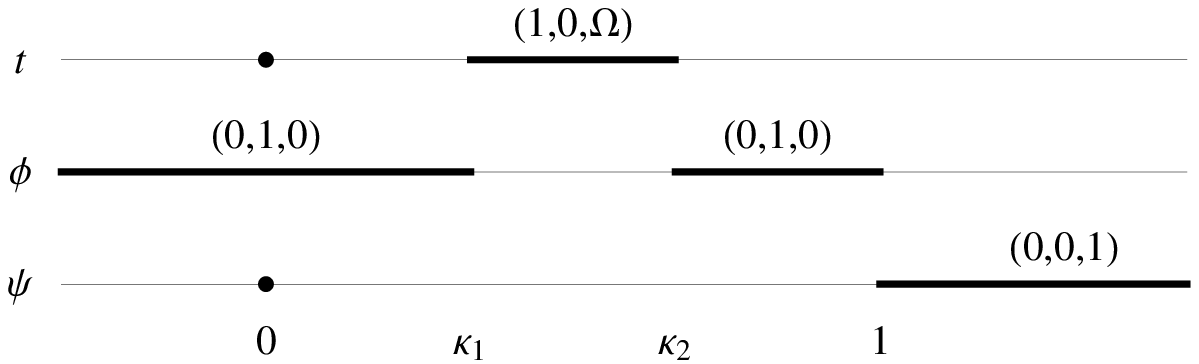}
\vspace{10pt}
\caption{Rod structure of the seed metric (left panel) and of the black ring solution, rotating along $S_1$ (right panel). The rods are located along the $z$-axis ($\bar{z}$-axis) with $\rho=0$ in the left (right) panel. The solid rods have positive density and the dashed rod has negative density.  The dots denote  ring singularities at $\bar{z} = 0$, which are removed by the fixing $b$ according to (\ref{bcond}).}  
\label{fig:rsBRS1}
\end{figure}

It was observed in \cite{Elvang:2007rd} that one can introduce rotation (along $S^1$) in the static black ring by performing a single soliton transformation; this is achieved by including a negative density rod on the rod structure of the seed metric to facilitate the addition of the angular momentum to the static black ring, when applying the inverse scattering method. 
The rod structures of the seed metric and of the resulting solution are shown in  Fig. \ref{fig:rsBRS1}. The solution is characterised by four dimensionful parameters: the length of the two finite rods, $i.e.$ $a_{32}$ and $a_{43}$ ($a_{ij}\equiv a_i-a_j$); the BZ ($i.e.$ Belinsky-Zakharov, from the inverse scattering method) parameter $b$; and the length $a_{21}$ of the phantom rod of the seed metric. However, it is useful to take out the overall scale of the solution and then introduce dimensionless parameters reflecting the length of the finite rods. We choose the overall scale $L$ to be 
$$L^2=a_4-a_1\ ,$$ 
and then introduce two dimensionless parameters $\kappa_i$ as 
$$\kappa_i=\frac{a_i-a_1}{L^2}\ , \,  \, \textrm{for} \, \, i=1,2.$$ 
Then we shifted the whole rod configuration along the $z$-axis, $i.e.$ $z \rightarrow \bar{z}+a_1$, which explains the labelling of the rod endpoints  in  Fig. \ref{fig:rsBRS1} (right panel).  
Due to the addition of the phantom rod, we are left with a singularity at $(\rho=0,\bar{z} = 0)$. This shows up as a $\bar{z}^{-1}$ divergence in the metric component $g_{\psi \psi}$ as indicated by the dots in Fig. \ref{fig:rsBRS1} (left panel). The singularity is, however, completely removed by setting
\
\begin{equation}
    b=\pm \sqrt{2 \kappa_1 \kappa_2} L \, .
\label{bcond}
\end{equation}
With this choice for $b$, the metric is completely smooth across $(\rho=0,\bar{z} = 0)$, and the final solution is fully characterised by $L, \kappa_1,$ and $\kappa_2$.  
The explicit line element can be found, in Weyl coordinates but in a different parameterisation, $e.g.$ in \cite{Harmark:2004rm}.

The ADM mass, ADM angular momentum, angular velocity of the horizon $\Omega_{\psi}$, Hawking temperature, $T_H$  and event horizon area, $A_H$, read:
\begin{equation}
\begin{array}{c}
    \displaystyle{M_{ADM}=\frac{3\pi}{4} L^2 \kappa_2\ , \qquad  J_{ADM}= - \frac{\pi}{2} L^2 b \ , \qquad   \Omega=-\frac{b}{2 L^2 \kappa_2} \ ,}  \\ \displaystyle{T_{H}=\frac{1}{2\pi} \frac{(1-\kappa_1)}{\sqrt{2 \kappa_2(\kappa_2-\kappa_1)}L }\ , \qquad  A_H= \frac{2 \pi (\kappa_2-\kappa_1)L^2}{T_H}\ .}
    \end{array} 
\end{equation}
A straightforward computation shows that the physical quantities computed above satisfy the Smarr relation \eqref{smarr}, if $S=A_H/4$ and 
if $b$ is fixed according to (\ref{bcond}).

Generically the solution contains a conical singularity. We choose it to be placed along the  finite space-like rod. Observe that it may either be a conical excess $\delta<0$, if the ring is under-spinning, or a conical deficit $\delta>0$, if it is over-spinning. The excess/deficit is given by
\begin{equation}
    \delta=2\pi \Bigg(1-\frac{1-\kappa_1}{\sqrt{1- \kappa_2}}\Bigg)\ .
\label{regcond}
\end{equation}

\begin{figure}
\centering
\includegraphics[width=0.45\textwidth]{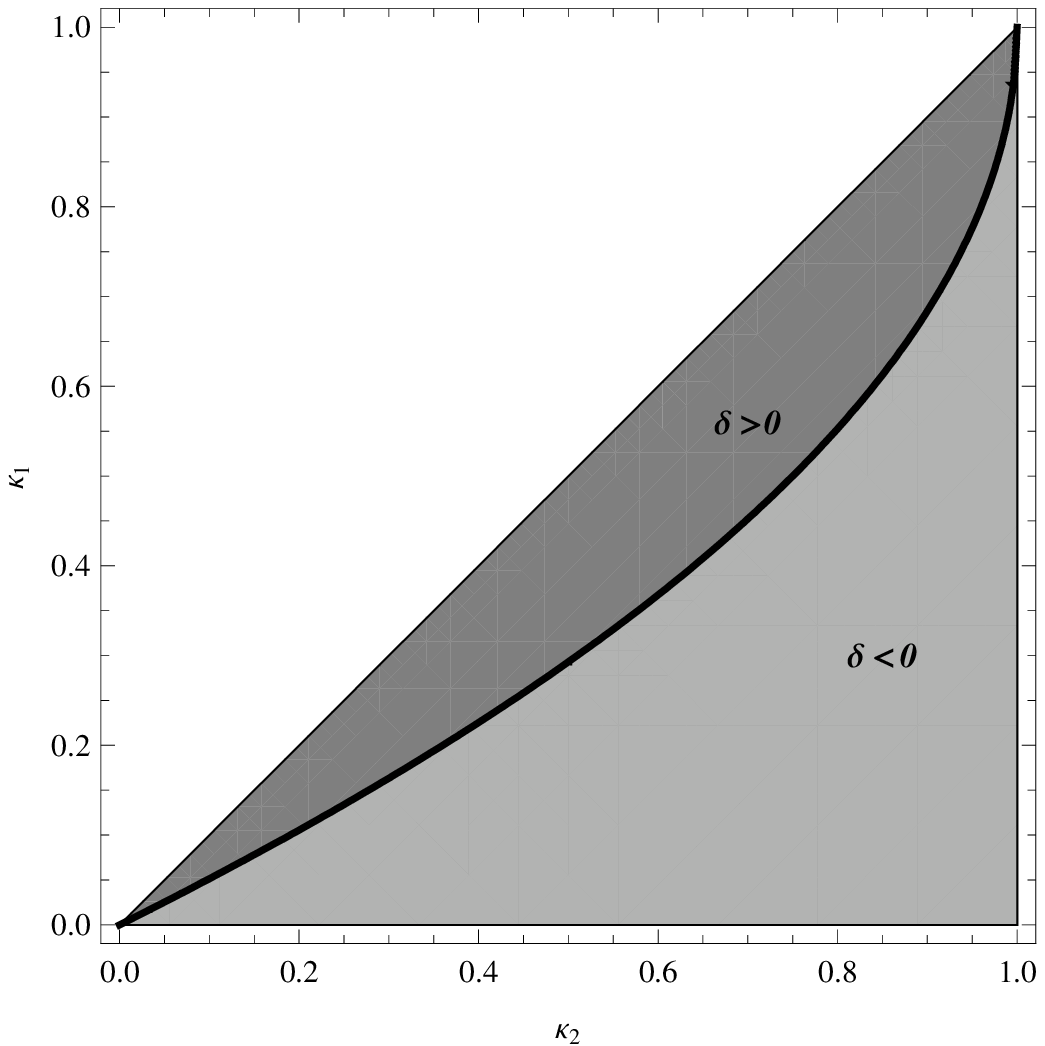} \ \ \ \ \ 
\includegraphics[width=0.45\textwidth]{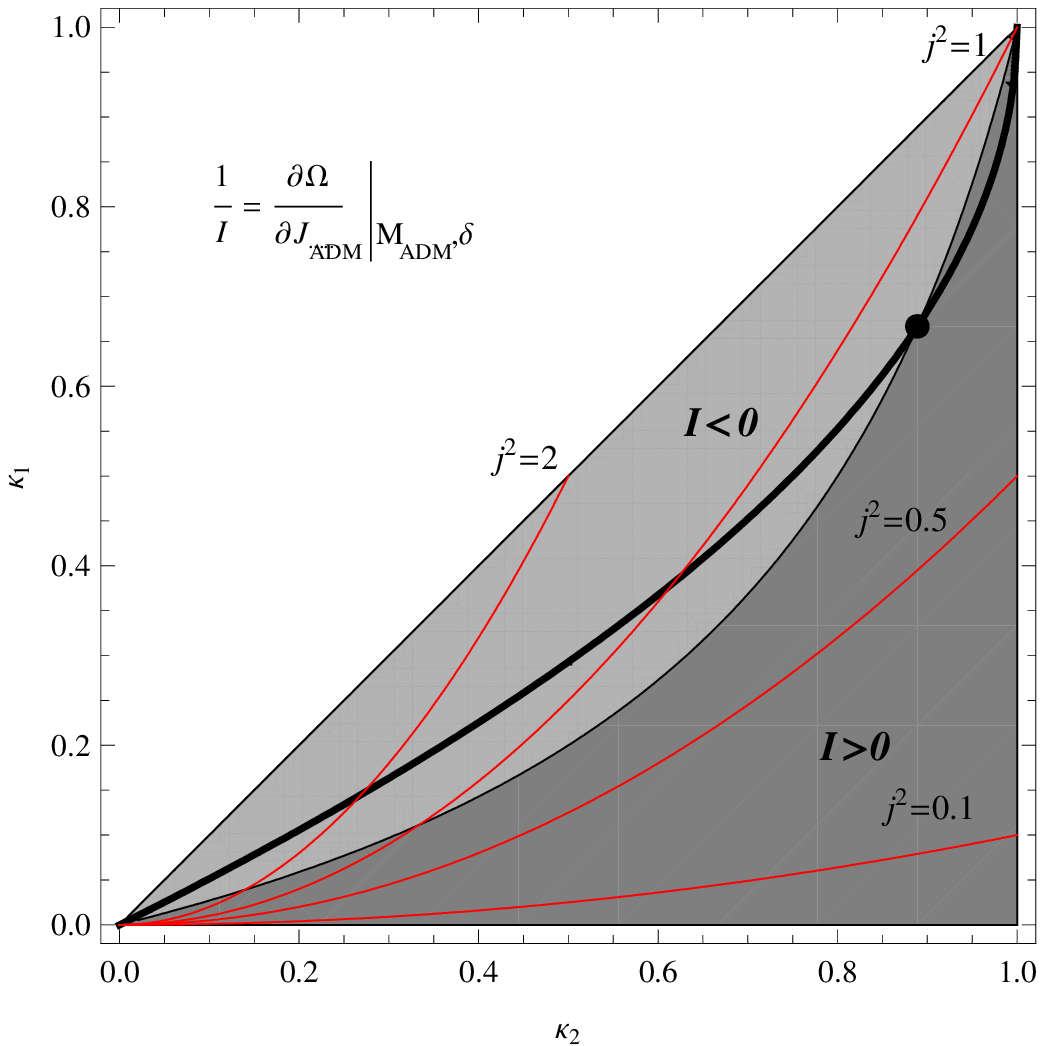}
\caption{Parameter space of $S^1$ rotating black ring solutions. The regular solutions are along the solid line. On either side of this line they are not in mechanical equilibrium, being held by a membrane-like conical deficit ($\delta>0$) or conical excess ($\delta<0$) (left panel). We also exhibit the sign of the mechanical moment of inertia of the solutions , together with lines of constant reduced spin (right panel). The dot marks the regular solution with $I=0$; $i.e.$ the meeting point between fat and thin ring branches.}  
\label{BRdelta}
\end{figure}

In Fig. \ref{BRdelta}, the sign of $\delta$ is displayed, together with the mechanical moment of inertia, in parameter space. This mechanical moment of inertia reports the variation of angular velocity, $\Omega$ with the angular momentum, $J_{ADM}$, keeping fixed $M_{ADM}$ and  $\delta$, $i.e.$
$$ \frac{1}{I}=\frac{\partial{\Omega}}{\partial{J_{ADM}}}\Bigg|_{M_{ADM},\delta}
=\frac{1-\kappa_1}{\pi L^4(-4\kappa_1+\kappa_2+3\kappa_1\kappa_2)} \, .$$

The sign of $I$ depends on $\kappa_1$ and $\kappa_2$, but not on $L$. Thus, it can be completely exhibited by a plot on the $\kappa_1-\kappa_2$ plane - Fig \ref{BRdelta}.  As usual, one can distinguish between the \textit{fat black ring branch} with $I>0$ from \textit{thin black ring branch} with $I<0$, regardless of the value of $\delta$.

Requiring regularity on the rod $\kappa_2<\bar{z}<1$, fixes the period of $\phi$ to be $\Delta \phi=2\pi$, which corresponds to $\delta=0$. From (\ref{regcond}), one can see that this balance is achieved whenever 
$$\kappa_2= \kappa_1(2-\kappa_1)\ .$$
In particular, for black rings in equilibrium, $I=0$ if $\kappa_1=2/3$, which leads to the familiar reduce spin value:
$$j^2=\frac{27 \pi}{32}\frac{J_{ADM}^2}{M_{ADM}^3}=\frac{\kappa_1}{\kappa_2^2}
=\frac{27}{32}, \,$$
where the thin and the fat balanced black rings meet.

The parameter $\mathcal{A}$, which is one of the relevant thermodynamical quantities, reads
\begin{equation}
    \mathcal{A}=\frac{Area}{\beta}=2\pi L^2 \frac{\;\;(1-\kappa_2)^{3/2}}{1-\kappa_1} \ ,
\end{equation} 
where $Area$ is the area of the surface spanned by the conical singularity, which is located at  $(\rho=0,\kappa_2<\bar{z}<1)$. The grand canonical potential, from \eqref{gibbs}, is simply
\begin{equation}
W=\frac{\pi}{4} L^2 \kappa_2+\frac{\pi}{2} L^2 (1-\kappa_2)\Bigg( 1-\frac{\sqrt{1-\kappa_2}}{1-\kappa_1}\Bigg)\ .
\end{equation}

\begin{figure}
\centering
\includegraphics[width=0.45\textwidth]{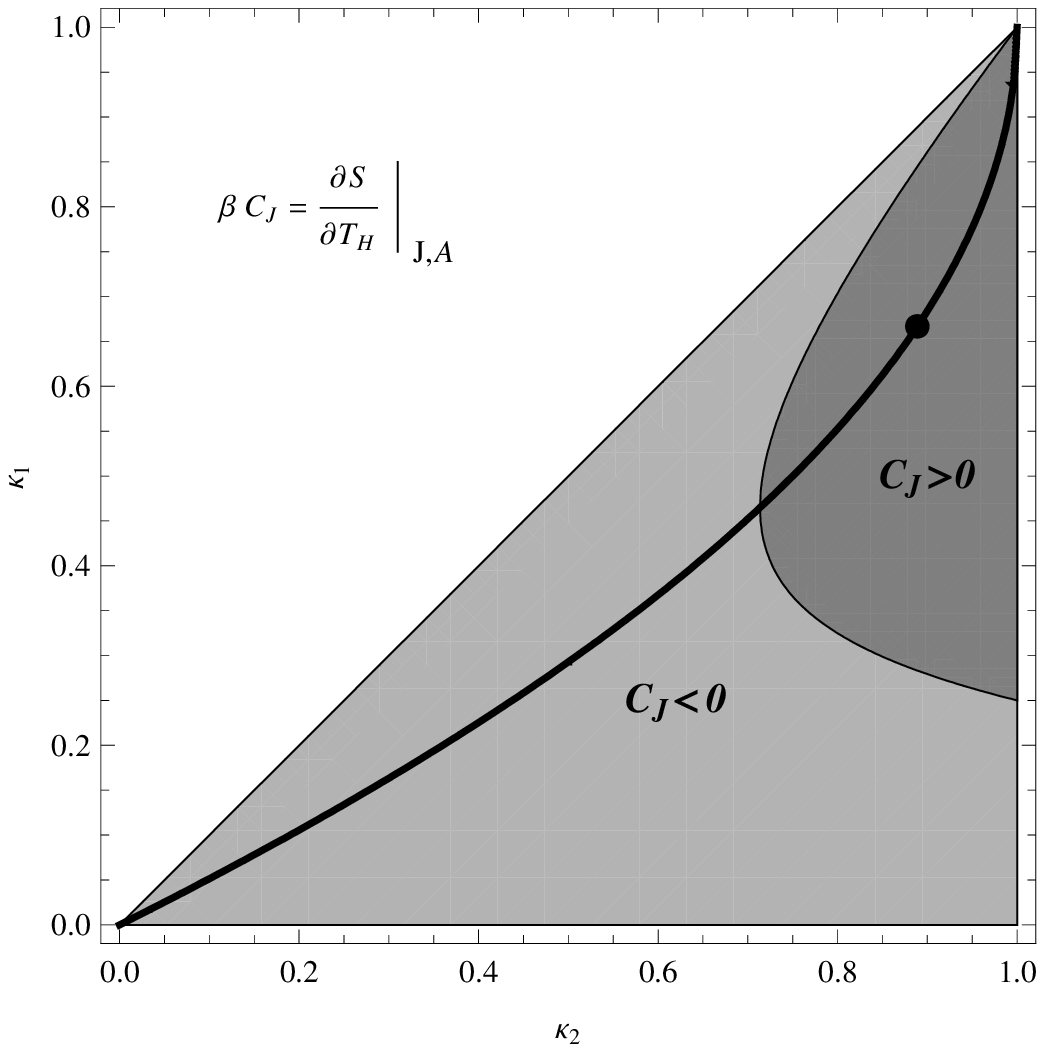} \ \ \ \ \ 
\includegraphics[width=0.45\textwidth]{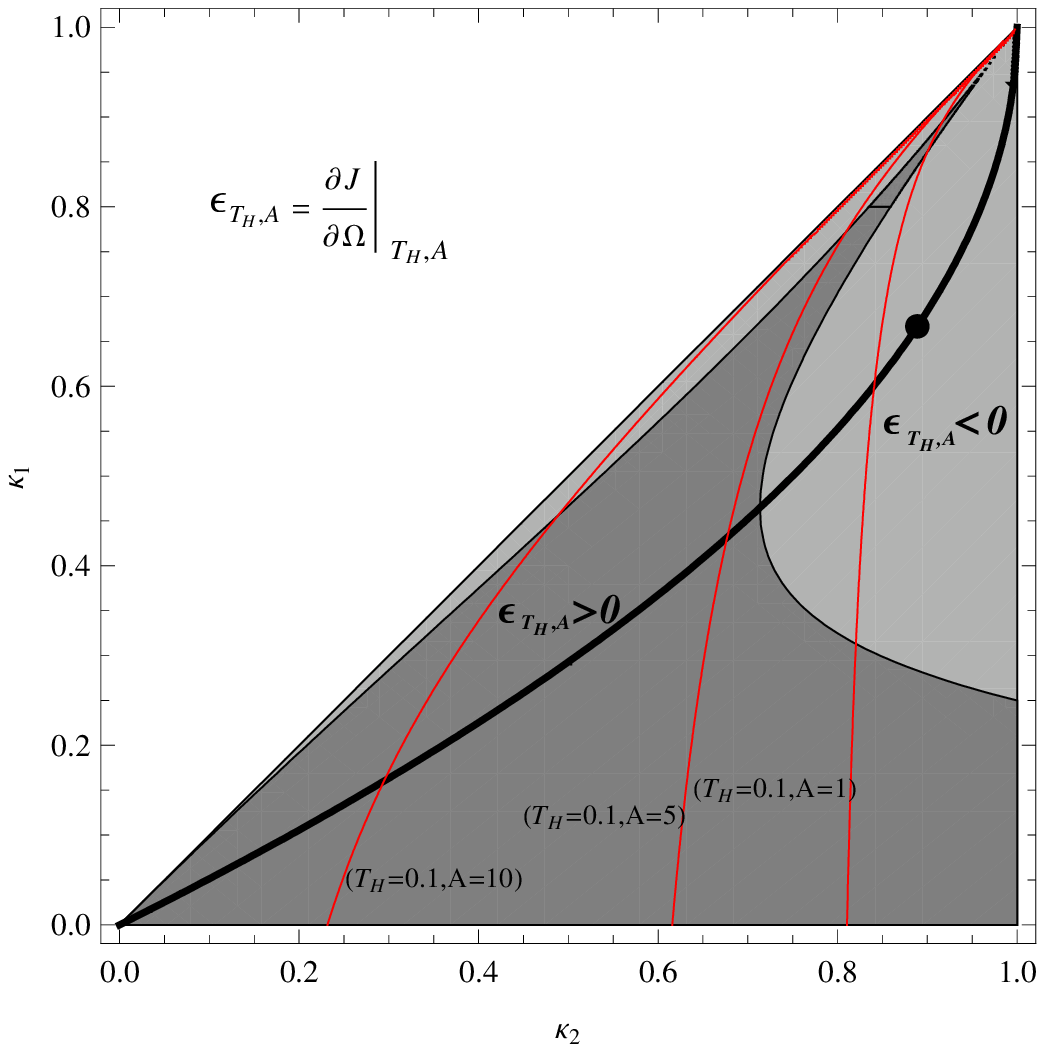}
\caption{Sign of: specific heat at constant $J,\mathcal{A}$, computed in the canonical ensemble (left panel); isothermal moment of inertia, at constant $\mathcal{A}$, computed in the grand canonical ensemble (right panel). Light (dark) grey regions correspond to positive (negative) sign. The solid line corresponds to regular black rings, wherein the ball separates the fat (below) from the thin (above) phases. Isotherms at constant $A$ are also displayed in the right panel.}  
\label{cjfig}
\end{figure}

It is not elegant to express in a simple way\footnote{For a balanced black ring, the
Gibbs free energy has the simple expression \cite{Astefanesei:2005ad}
\begin{equation}
W=\frac{1}{16 \pi T_H^2}\left(-1+\sqrt{\frac{16\pi^2 T_H^2}{\Omega^2}-1}\right).
\end{equation}} the parameters $L$,$\,\kappa_1$, and $\kappa_2$  and thus $W$
in terms of $T_H$, $\Omega$ and $\mathcal{A}$. Nevertheless, when this last three quantities are used as thermodynamical variables, we precisely recover the Bekenstein-Hawking area law for the entropy:
\begin{eqnarray}
    S &=&  -\frac{\partial{W}}{\partial{T_H}}\Bigg|_{\Omega,\mathcal{A}} 
= \pi^2 L^3 \frac{\kappa_2-\kappa_1}{1-\kappa_1}\sqrt{2 \kappa_2 (\kappa_2-\kappa_1)}
= \frac{A_H}{4} \ .
\end{eqnarray}
Moreover, the thermodynamical angular momentum yields the ADM angular momentum and the pressure gives the expected relation \eqref{pressure}:
\begin{equation}
   J=- \frac{\partial{W}}{\partial{\Omega}}\Bigg|_{T_{H},\mathcal{A}}=J_{ADM}\ , \qquad  P=\frac{\partial{W}}{\partial{\mathcal{A}}}\Bigg|_{T_{H},\Omega}=-\frac{\delta}{8\pi} \ . \label{jpring}
\end{equation}
Finally, observe that the thermodynamical mass $\mathcal{M}$, computed from $\mathcal{M}=W+T_H S+\Omega J$, yields exactly \eqref{thermomass}, upon using \eqref{bcond} -- \eqref{jpring}.

\begin{figure}
\centering
\includegraphics[width=0.45\textwidth]{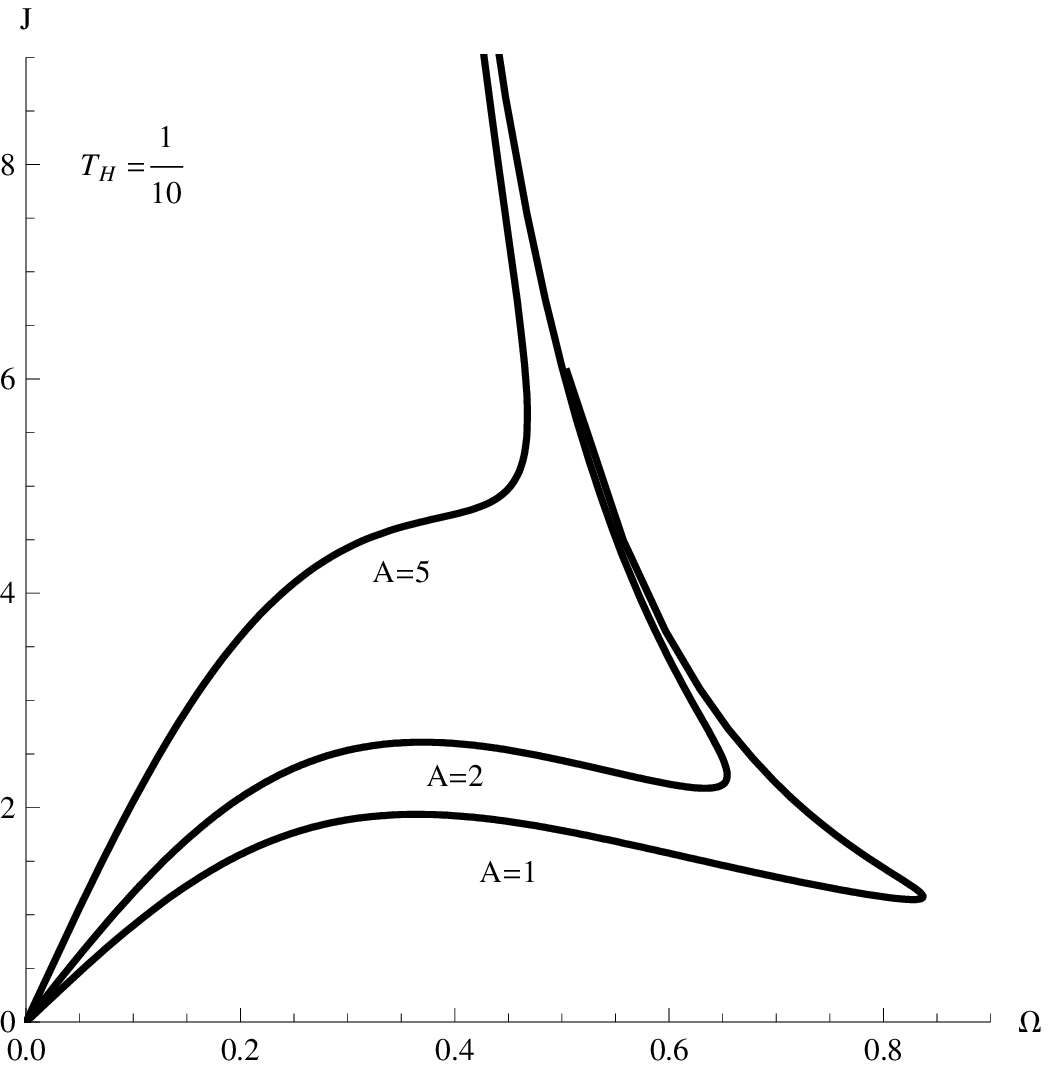} \ \ \ \ \ 
\includegraphics[width=0.45\textwidth]{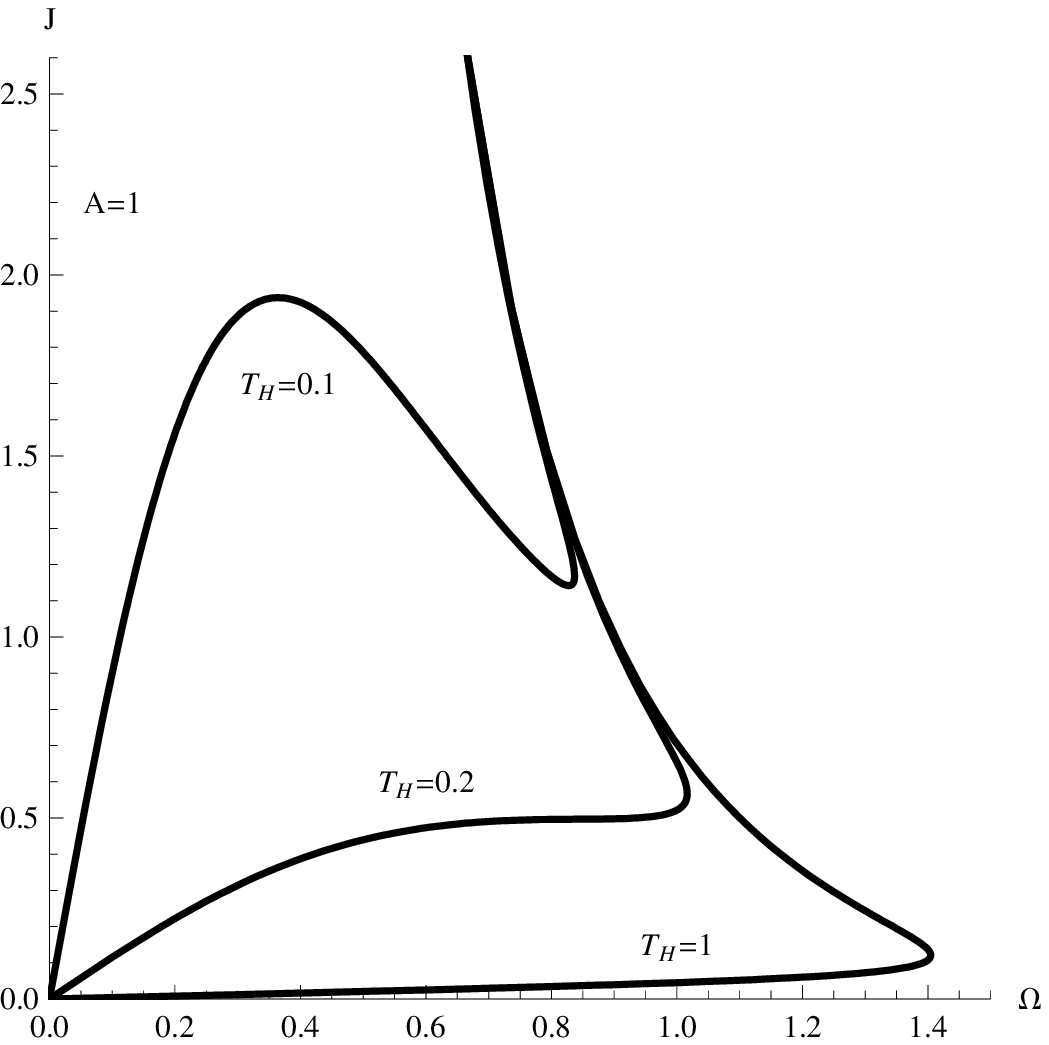}
\caption{Isotherms - at constant $T_H,\mathcal{A}$ - for the $S^1$ rotating black ring. Each $(J_{ADM},T_H,\mathcal{A})$ generally corresponds to three values of $\Omega$ for low values of $T_H$ and $\mathcal{A}$. For high values of $T_H$ and/or $\mathcal{A}$, $\epsilon_{T_H,\mathcal{A}}$ just changes sign once.}  
\label{fig:IsoBR}
\end{figure}

The thermodynamical stability of the $S^1$ spinning black ring is analysed in Fig. \ref{cjfig}, where the signs of the isothermal moment of inertia, 
$$ \epsilon_{T_H,\mathcal{A}}= \pi L^4 \kappa_2 \frac{
(4 - \kappa_2) \kappa_2 + 
   4 \kappa_1^2 (2 + \kappa_2) - 
   3 \kappa_1 \kappa_2 (4 + \kappa_2)}{(4 - \kappa_2) \kappa_2 + \kappa_1 (-4 + \kappa_2^2)} \, \ ,$$
 and specific heat at constant angular momentum,
$$C_J= \pi^2 L^3 \sqrt{2 \kappa_2}  \,
 \frac{(\kappa_2-\kappa_1)^{3/2}(\kappa_2 (8 + \kappa_2) +\kappa_1 (4 - 16\kappa_2 + 
    3\kappa_2^2))}{(\kappa_1-1)(
(4 - \kappa_2) \kappa_2 + 
   4 \kappa_1^2 (2 + \kappa_2) - 
   3 \kappa_1 \kappa_2 (4 + \kappa_2))} \, \ ,$$
are plotted. The plots are in the $\kappa_1$, $\kappa_2$ space; indeed, as may be checked in the last two formulae, the $L$ dependence does not change the sign of these quantities. It can be observed that there is no region in parameter space wherein both quantities are positive, and hence no region wherein the solution is thermodynamically stable in the grand canonical ensemble. This situation is analogous to that of the Kerr solution in four dimensions  \cite{Monteiro:2009tc}. In Fig. \ref{fig:IsoBR}, isotherms for various values of $A$ are plotted in the $J$-$\Omega$ plane; the regions with negative and positive isothermal moment of inertia are clear and in correspondence with those plotted in Fig. \ref{cjfig}.

\subsection{$S^2$ rotating black ring} 

This solution is explicitly given in \cite{Figueras:2005zp}, where all relevant  physical quantities where calculated. The metric is written in $(t,x,y,\phi,\psi)$ coordinates, where $\psi$ parameterises an $S^1$ and $(x,\phi)$  an $S^2$. This particular black ring is rotating along the azimuthal direction $\phi$ of the $S^2$. The coordinates $x$ and $y$ vary within the ranges $-1 \le x \le 1,$ $-\infty < y \le -1,$ with asymptotic infinity at $x,y=-1$. 

After requiring regularity at infinity, the physical parameters for the black ring mass, angular momentum, temperature, angular velocity and horizon area are
\begin{equation}
\begin{array}{c}
     \displaystyle{ M_{ADM} =  \frac{3\pi R^2\lambda}{4(1-\lambda+\nu^2)} \ , \ \  
      J_{ADM} = - \frac{ \pi R^3\lambda \nu}{(1-\lambda+\nu^2)^{3/2}}\ , \ \ \Omega=-\frac{\sqrt{1-\lambda+\nu^2}(\lambda-\sqrt{\lambda^2-4\nu^2})}
{2 \lambda \nu R}\ , } 
\\
 \displaystyle{     T_H = \frac{\sqrt{(\lambda^2-4\nu^2)(y_h^2-1)}}{4\pi R \lambda}\ , \qquad 
A_H = \frac{8 \pi^2 R^3 \lambda}{(1-\lambda+\nu^2)\sqrt{y_h^2-1}} \ ,}
\end{array}
\end{equation}
where $$y_h=\frac{-\lambda+\sqrt{\lambda^2-4\nu^2}}{2\nu^2}$$ determines the location of the horizon. 
The parameters $\lambda$ and $\nu$ take values $2\nu<\lambda<1+\nu^2$. In the limit $\nu \rightarrow 0$, we recover the static black ring.

The solution contains a conical excess angle $\delta$ along $x=1$ and 
$y_h \le y \le -1$: 
\begin{equation}
    \delta=2\pi \Bigg(1-\sqrt{\frac{1+\lambda+\nu}{1-\lambda+\nu}}\Bigg).
\end{equation}
One can easily see that it is impossible to required regularity for nonzero $\lambda$, 
which means that a $S^2$ rotating black rings cannot be in equilibrium \cite{Figueras:2005zp}. Computing the mechanical moment of inertia
$$ \frac{1}{I}=\frac{\partial{\Omega}}{\partial{J_{ADM}}}\Bigg|_{M_{ADM},\delta}
=\frac{(1 - \lambda + \nu^2)^2 (\lambda^2 - 
   4 \nu^4 - \lambda \sqrt{\lambda^2 - 4 \nu^2})}{2 \pi R^4 \lambda^2 \nu^2 \sqrt{\lambda^2 - 4 \nu^2}} \, ,$$
it can be observed that just like for ordinary black holes, $I$ is always positive - increasing $J_{ADM}$, $\Omega$ also increases.

The parameter $\mathcal{A}$, related with the  area of the surface spanned by the conical singularity, is:
\begin{eqnarray}
    \mathcal{A}=\frac{Area}{\beta} =
 \pi R^2 \sqrt{\frac{1+\lambda+\nu}{1-\lambda+\nu}}\,
\frac{2-\lambda-\sqrt{\lambda^2-4\nu^2}}{2+\lambda+\sqrt{\lambda^2-4\nu^2}} \ . 
\end{eqnarray}

The Gibbs free energy is given by (\ref{gibbs}). From it, we precisely recover, using \eqref{thermorelation}, $S=A_H/4$, $J=J_{ADM}$ and $P=-\delta/8\pi$.

\begin{figure}
\centering
\includegraphics[width=0.45\textwidth]{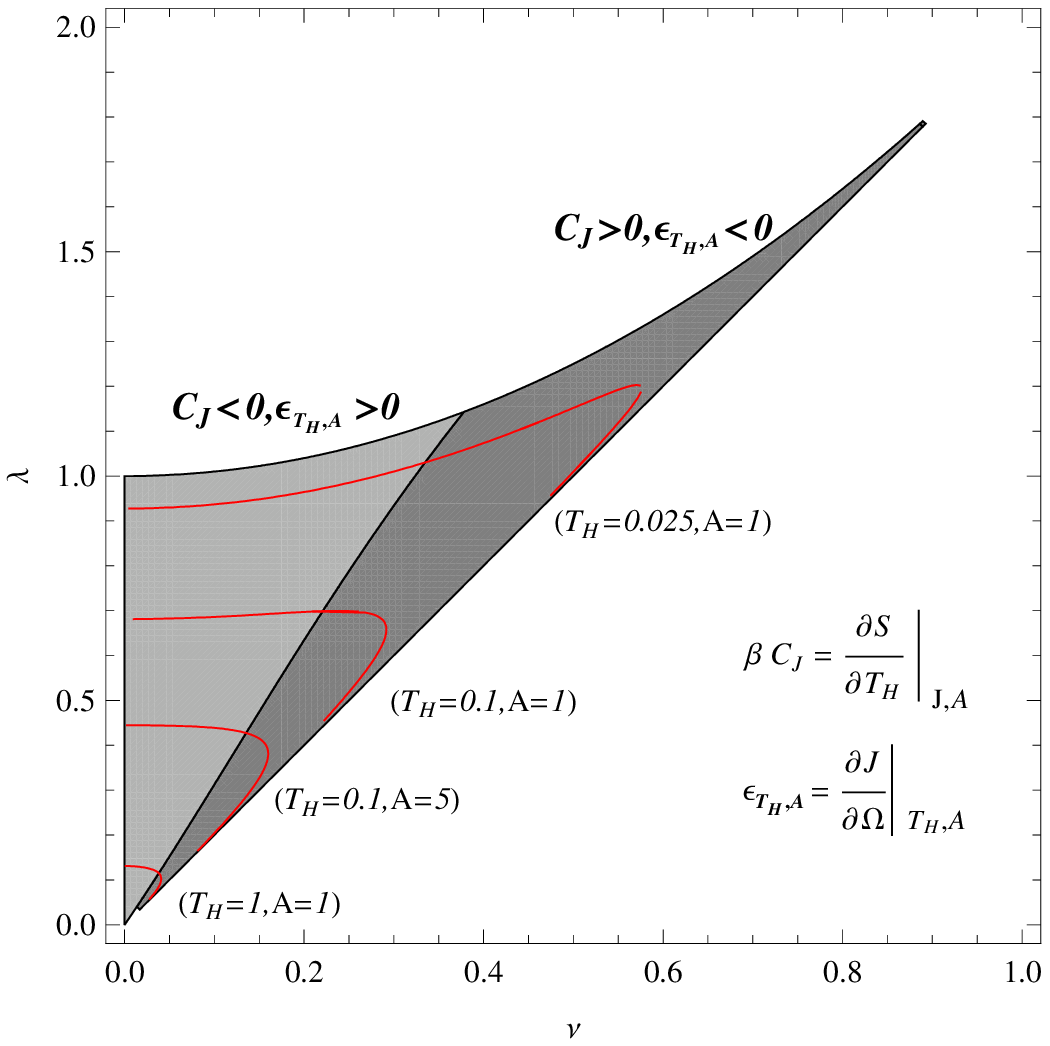} \ \ \ \ \ \ \ \ 
\includegraphics[width=0.4\textwidth]{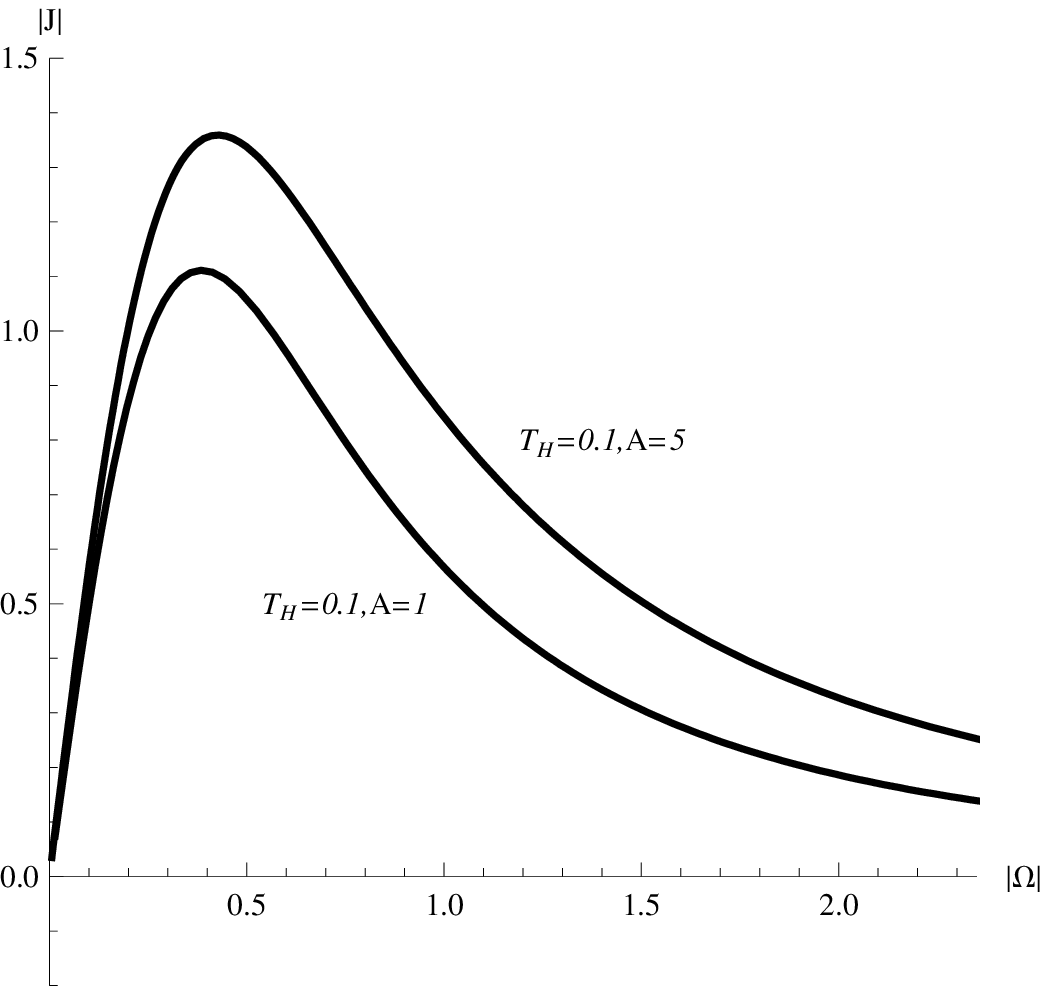}
\caption{Sign of the isothermal moment of inertia, 
$ \epsilon_{T_H,\mathcal{A}},$ and specific heat at constant angular momentum,
$C_J$, in the parameters space of  $S^2$ rotating black ring (left panel). Isotherms in the $J$-$\Omega$ plane, for various values of $A$ (right panel).}  
\label{fig:S2ring}
\end{figure}

The thermodynamical stability of the $S^2$ rotating black ring is analysed in Fig. \ref{fig:S2ring}, where the signs of the isothermal moment of inertia, 
$ \epsilon_{T_H,\mathcal{A}},$ and specific heat at constant angular momentum,
$C_J,$ are plotted. For any values of the parameters, when $ \epsilon_{T_H,\mathcal{A}}$ is positive,  $C_J$ is negative and vice-versa. Therefore, the $S^2$ rotating black ring is also thermodynamically unstable in the grand-canonical ensemble.

\section{The co-rotating double-Kerr solution} 
The approach proposed in this work can be extended to multi-black objects as well. 
For solutions with a non-connected event horizon, one needs to require that all connected components 
possess the same temperature and angular velocity, to ensure thermodynamical equilibrium. 
Interestingly, these conditions do not impose  mechanical equilibrium
and one can discuss the thermodynamics of multi-black holes with conical singularities.

Perhaps the most obvious example here corresponds to the asymptotically flat $d=4$ double Kerr solution, with two black holes having the same Komar mass and angular momentum. 
We further impose the axis condition  \cite{Herdeiro:2008kq,Costa:2009wj}, which guarantees that the ADM mass (angular momentum) is the sum of the Komar masses (angular momenta) of the individual black holes. The solution is then characterised by three physical quantities corresponding to the mass $M$ and  the angular momentum $J$ of each black hole and the distance between them. 
\begin{figure}
\centering
\includegraphics[width=0.6\textwidth]{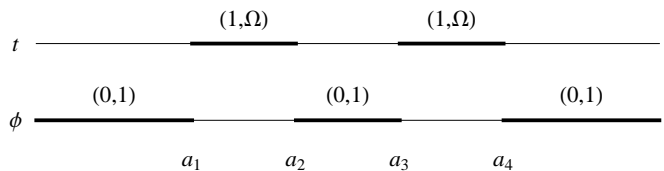}
\caption{Rod structure of the asymptotically flat, equal mass and angular momentum double-Kerr solution, obeying the axis condition.} 
\label{fig:rsDK}
\end{figure}

The simplest way to parameterise the solution and the physical quantities is in terms of the two BZ parameters $b$ and $c$, the length  $a_{21}=a_{43}=\sigma$ of the rods associated to the black hole horizon and the length $a_{32}=\lambda$ of the rods between the black holes. The rod structure is represented in Fig. \ref{fig:rsDK}. Additionally, it is necessary to consider the following constraint \cite{Costa:2009wj}
\begin{equation}
    (1+2c^2+b^2 c^2)\left(\frac{\sigma}{\lambda+\sigma}\right)^2+2c \frac{1-b^2c^2}{b-c}\frac{\sigma}{\lambda+\sigma}=(1-bc)^2 \ ,
\label{axiscondition}
\end{equation}
which guarantees that the axis condition is obeyed. The line element of the resulting configuration can be constructed from the data presented in \cite{Herdeiro:2008kq,Costa:2009wj}.

The physical quantities are given by:
\begin{eqnarray}
M_1=M_2 &=& \sigma \frac{(1 + b^2) (1 - b c) (\lambda + \sigma) (\lambda + 
   2 \sigma) }{2 ((1 - b c) \lambda + 
   2 \sigma)} \frac{\eta}{ \Delta}\ , \\
J_1=J_2 &=& \sigma M  \Delta \, \eta \frac{ (b (b c-1) \lambda + ((1 + b^2) c - 2 b) \sigma) }{ (1 - b c) ((1 - b c)^2\lambda
( \lambda+   2  \sigma) + (1 - b^2) (1 - 
      c^2) \sigma^2)^2}\ ,  \\
\Omega &=& \frac{\Delta}{\eta^2}  \frac{((b^2 - 1 ) c \lambda (\lambda + 2 \sigma) + 
  b (c^2 - 1) (\lambda^2 + 2 \lambda \sigma + 2 \sigma^2))}{2 \sigma (1 + b^2)  (\lambda^2 + 3 \lambda \sigma + 
   2 \sigma^2)} \ , \\
T_{H} &=& \frac{\Delta^2 }{4  \pi \sigma \, \eta^2 (1 + b^2) (\lambda^2 + 3 \lambda \sigma + 
   2 \sigma^2)} \ , \\
A^1_{H}=A^2_{H}  &=& \frac{\sigma}{T_H}  \ , 
\end{eqnarray}
where the individual masses and angular momenta are computed as Komar integrals and
\bequ  \Delta \equiv  ((1 - b c) \lambda + (1 + b) (1 - c) \sigma) ((1 - b c) \lambda + (1 - b) (1 + c) \sigma) \ , \qquad  \eta \equiv  (1- b c) \lambda+(1 - c^2) \sigma \ . \eequ 
It is now easy to check that once again the physical quantities satisfy the Smarr relation \eqref{smarr}, taking $M_{ADM}=M_1+M_2$, the total angular momentum to be $J_{ADM}=J_1+J_2$ and the entropy to be $(A^1_H+A^2_H)/4$.

As argued by several authors ~\cite{dietz,Manko2000,manko2001,Neugebauer:2009su}, the spin-spin repulsion cannot balance the gravitational attraction in a double Kerr system where both Kerr objects are black holes. In this particular co-rotating limit one has the following excess angle along the section in between the black holes:
\begin{equation}
    \frac{\delta}{2\pi}=-\frac{(1 + 4 b c - c^2 - b^2 (1 - c^2)) \sigma^2}{(1 - b c)^2 \lambda^2 + 2 (1 - b c)^2 \lambda \sigma - 
 4 b c \sigma^2}\ .
\end{equation}

The parameter $\mathcal{A}$ associated to this conical singularity is
\begin{equation}
    \mathcal{A}= \frac{\lambda}{\Delta} ((1 - b c) \lambda + 2 \sigma) ((1 - b c) \lambda -   2 b c \sigma)\ .
\end{equation}

Although simple to calculate, the free energy of this co-rotating system, given by \eqref{gibbs}
is already a quite long expression. A careful analysis, where the constraint (\ref{axiscondition}) is cautiously considered, shows (numerically) the expected results.
\bequ
    S=-\frac{\partial{W}}{\partial{T_H}}\Bigg|_{\Omega,\mathcal{A}} = \frac{A^1_H+A^2_H}{4} \ , \qquad J=-\frac{\partial{W}}{\partial{\Omega}}\Bigg|_{T_{H} ,\mathcal{A}} =  J_1+J_2 \ , \qquad
P=\frac{\partial{W}}{\partial{\mathcal{A}}}\Bigg|_{T_{H},\Omega}=-\frac{\delta}{8\pi}\ . \eequ

Note that this perfect match between the entropy and the sum of horizon areas, also between the different angular momenta, is not possible if instead of
(\ref{firstgibbs}) we consider that the variation of the free energy $W$ 
is given by  $  dW = -S dT_{H} - \mathcal{J}d\Omega -\mathcal{F}d\lambda$ as in \cite{Costa:2009wj}.

\begin{figure}
\centering
\includegraphics[width=0.5\textwidth]{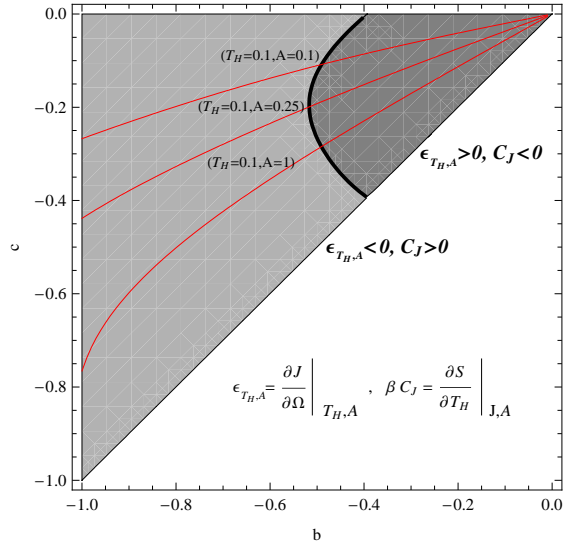} 
\caption{Sign of the isothermal moment of inertia, 
$ \epsilon_{T_H,\mathcal{A}},$ and specific heat, $C_J$, in a window of the $(b,c)$ parameters space for the co-rotating double-Kerr solution in thermodynamical equilibrium. Some curves of fixed $(T_H,\mathcal{A})$ are also displayed. } 
\label{fig:DKtherm}
\end{figure}
The thermodynamical stability of the co-rotating double Kerr system is analysed in Fig. \ref{fig:DKtherm}, where the signs of the isothermal moment of inertia and specific heat at constant angular momentum are plotted, for a particular window of the parameter space leading to physical solutions, $i.e.$ with positive masses and horizon areas. Since $C_J$ is proportional to $\lambda^2$, $\epsilon_{T_H,\mathcal{A}}$ is proportional to $\lambda^3$ and the length $\lambda$ is always positive, the sign of the previous function just depends on $(b,c)$, which explains the parameters space in Fig. \ref{fig:DKtherm}. Again, as for the single Kerr and the single rotating rings above, it can be observed that there is no region in parameter space wherein both quantities are positive, and hence no region wherein the solution is thermodynamically stable in the grand canonical ensemble. The isotherm behaviour, namely the variation of the angular moment with $(\Omega,\mathcal{A})$ for fixed $T_H$ is in complete agreement with what is described in  \cite{Costa:2009wj}. In particular the curve $J=J(\Omega)$ is just like the one describing a single Kerr back hole, when the two black holes are far away or infinitesimally close.

\section{The Black Saturn Solution}

The black Saturn \cite{Elvang:2007rd} is a $d=5$ asymptotically flat solution describing a black ring around a concentric Myers-Perry black hole.  
Both objects have angular momentum only in a single plane - on the plane of the ring along the $S^1$ direction. 
The solution was generated and thoroughly analysed in \cite{Elvang:2007rd}, requiring regularity (on and outside the even horizons). 

\begin{figure}
\centering
\includegraphics[width=0.6\textwidth]{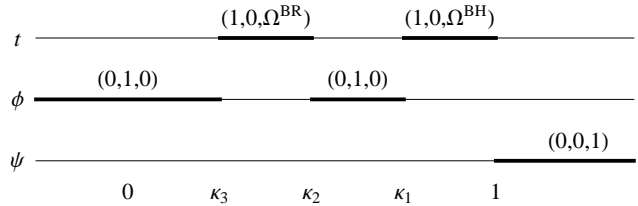}
\caption{Rod structure of the black Saturn solution.} 
\label{rsBS}
\end{figure}

We are now interested in the particular limit where the solution is in thermodynamical equilibrium, $i.e.$ when the black ring and the black hole have the same temperature and the same angular velocity, but not necessarily in mechanical equilibrium. 
The physical quantities of the general solution are explicitly given in \cite{Elvang:2007rd} in terms of five parameters: $L,\kappa_1,\kappa_2,\kappa_3$ and $\bar{c}_2$, where the last is a dimensionless parameter related to the original BZ parameter, $L$ is the overall scale and the three dimensionless parameters $\kappa_i$ are related with the rod endpoints - Fig. \ref{rsBS}. The physical quantities are explicitly presented in terms of this parameterisation in sec. 3.5 and 3.7 of \cite{Elvang:2007rd}.
In what follows, we denote the black hole (black ring) mass, angular momentum, horizon angular velocity, temperature and horizon area by $M^{BH}, J^{BH}, \Omega^{BH}, T_H^{BH}, A_H^{BH}$ ($M^{BR}, J^{BR}, \Omega^{BR}, T_H^{BR}, A_H^{BR}$). Observe that $M_{ADM}=M^{BH}+M^{BR}$ and $J_{ADM}=J^{BH}+J^{BR}$.

Setting $\Omega^{BH}=\Omega^{BR}$ and solving for $\bar{c}_2$ implies
\begin{equation}
   \bar{c}_2= - \frac{\kappa_3}{\kappa_1 \kappa_2}\frac{1-\kappa_1 }{1-\kappa_3}\ .
\label{equalav}
\end{equation}
Considering the previous constraint and demanding $T^{BH}_H=T^{BR}_H$ leads us to define $\kappa_3$ in terms of $\kappa_1$ and $\kappa_2$ through the relation:
\begin{equation}
   \kappa_3= \kappa_1 + \kappa_2 \,\frac{1 -\kappa_2}{1-\kappa_1-\kappa_2}\ .
\label{equalt}
\end{equation}
It is important to note that the $\kappa_i$'s satisfy the ordering $0 \leq \kappa_3 \leq \kappa_2 < \kappa_1 \leq 1$, and we can also add that (at least) for all physical solutions in thermodynamical equilibrium, $\kappa_1+\kappa_2 \geq 1$.

\begin{figure}
\centering
\includegraphics[width=0.4\textwidth]{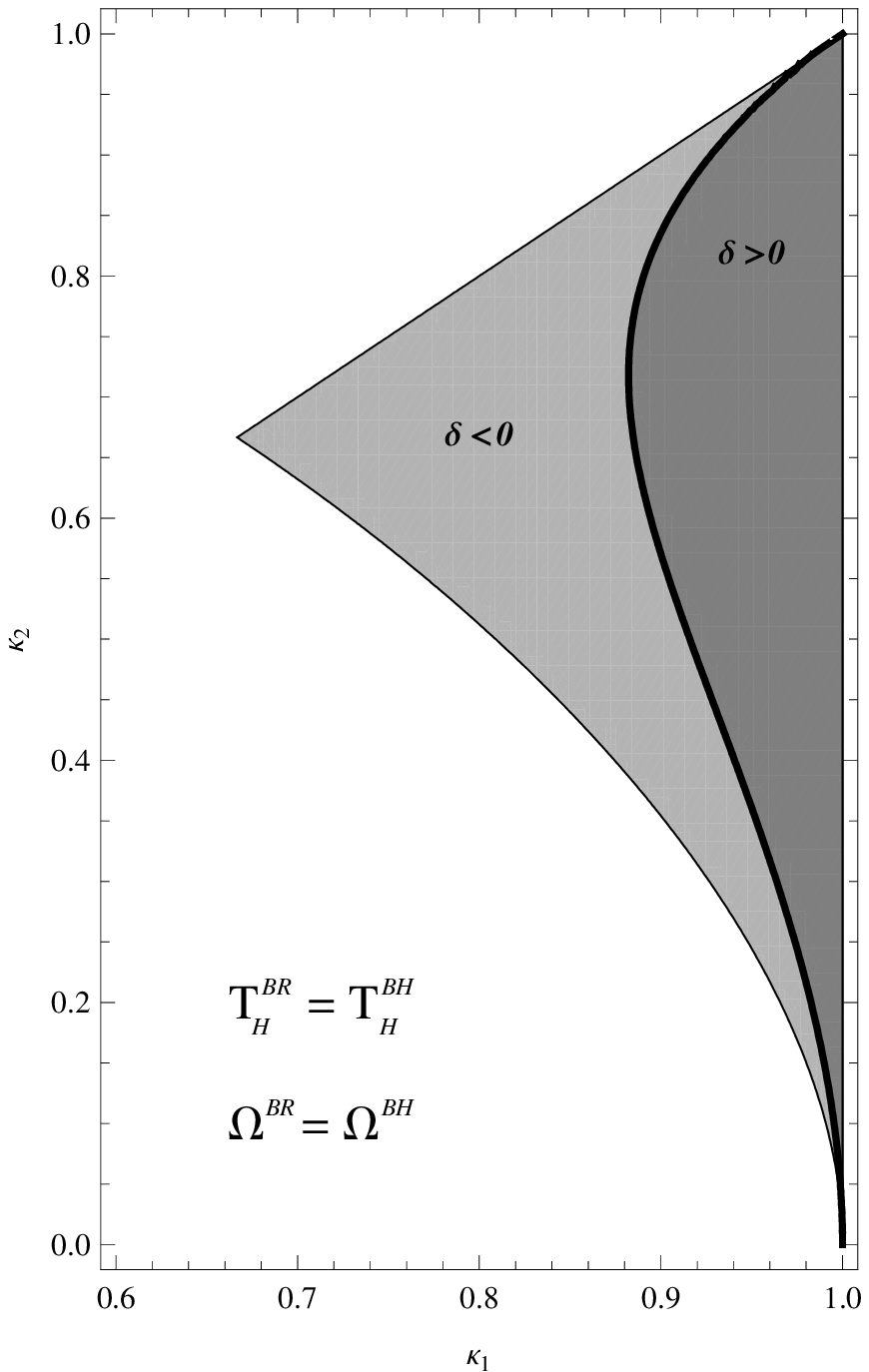} \ \ \ \ \ \ \ \ \ \ 
\includegraphics[width=0.4\textwidth]{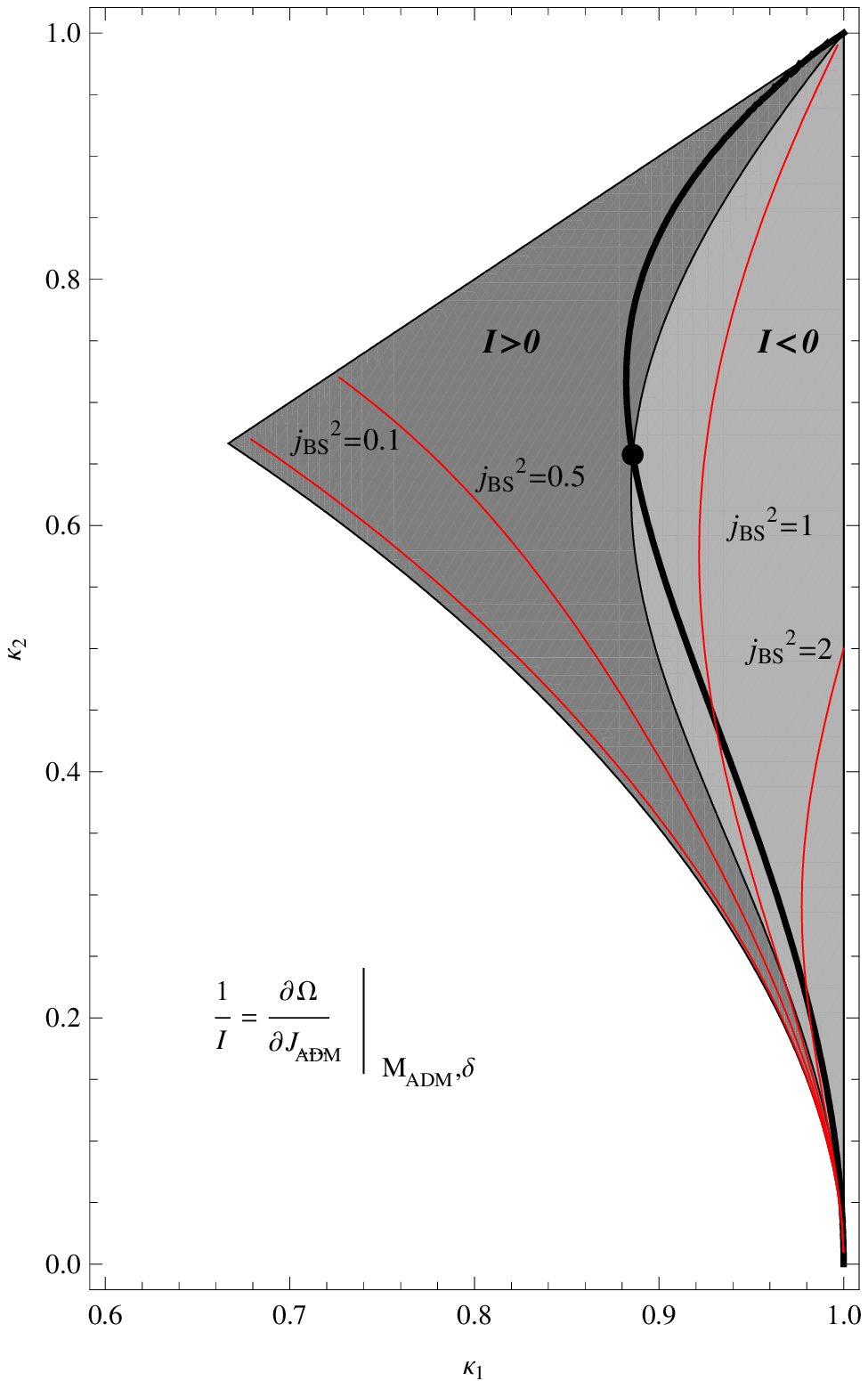}
\caption{The coloured region represents the space of black Saturn solutions in thermal equilibrium. The regular solutions are along the solid line. On either side of this line they are not in mechanical equilibrium, being held by a membrane-like conical deficit ($\delta>0$) or conical excess ($\delta<0$) (left panel). We also exhibit the sign of the mechanical moment of inertia of the solutions and some lines of constant reduced spin (right panel).} 
\label{fig:bsequilibrium}
\end{figure}

The original solution has a conical singularity membrane in the plane of the ring, extending from the inner $S^1$ radius of the black ring to the horizon of the $S^3$ black hole. Applying the constraints (\ref{equalav}) and (\ref{equalt}) that lead to equilibrium thermodynamics, the corresponding deficit angle is  
$$ \frac{\delta}{2\pi}=1-\frac{\kappa_2(1-\kappa_2)^2}
{(\kappa_1-\kappa_2)(\kappa_1+\kappa_2-1)}\sqrt{\frac{\kappa_2}{\kappa_1(\kappa_1 \kappa_2 -(\kappa_1+\kappa_2-1)^2)}} \ ,$$
where we can see that it is still possible to reach mechanical equilibrium, taking into account the properties of $\kappa_i$'s.  In Fig. \ref{fig:bsequilibrium} the sign of $\delta$ is analysed along the parameter space of black Saturn solutions in thermodynamical equilibrium. The balanced solutions lie along the solid line and have already been study in \cite{Elvang:2007hg}. This solutions can also be classify as $fat$ or $thin$ in function of the sign of the mechanical moment of inertia, $I$, just like single black rings.

The sign of $I$ is displayed in right panel of Fig. \ref{fig:bsequilibrium}, just in function of $(\kappa_1,\kappa_2)$ because $I$ is proportional to $L^4$ and so $L$ does not influence its sign. Analysing, along the parameter space, the family of solutions with constant reduced spin 
$$j_{BS}^2=\frac{27 \pi}{32}\frac{(J^{BR}+J^{BH})^2}{(M^{BR}+M^{BH})^3} \, ,$$ 
one can observed that, as expected, $\delta<0$ for under-spinning back Saturn configurations, while for over-spinning ones,  $\delta>0$. 
Generically, most of the angular momentum is in the ring. In the \textit{fat branch}, which is associate with lower reduce spin values, the fraction of the total mass that goes into the central black hole increases with  $\kappa_2$. For high $j_{BS}$, the ring also carries most of the total mass.

\begin{figure}
\centering
\includegraphics[width=0.31\textwidth]{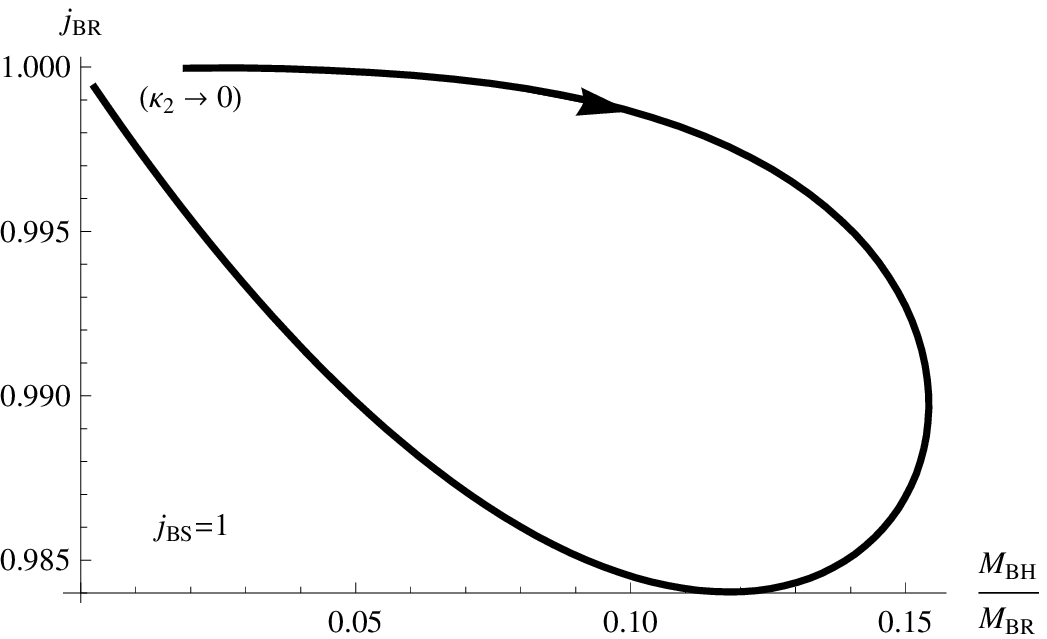} \ \
\includegraphics[width=0.31\textwidth]{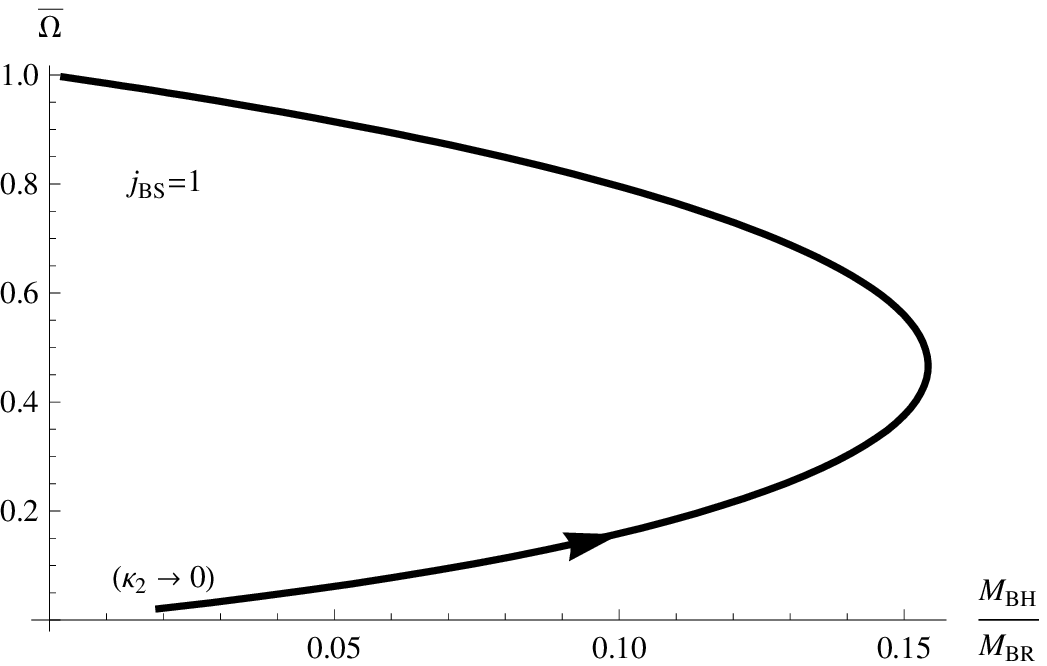} \ \  
\includegraphics[width=0.31\textwidth]{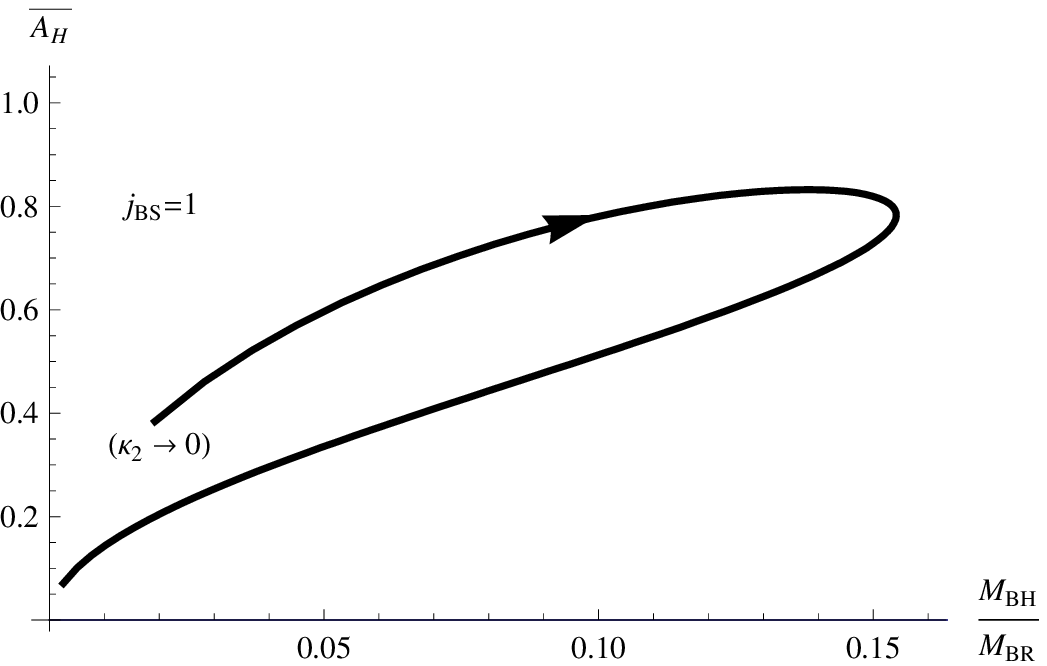}
\caption{Analysis of the black Saturn solution in thermodynamical equilibrium for fixed ADM mass and reduce spin $j_{BS}=1$. The plots exhibit how the relative mass $M^{BH}/M^{BR}$ varies with the reduce spin of the black ring $j_{BR}=\sqrt{{27\pi/32}}J^{BR}/M_{ADM}^{3/2}$ (left panel), the reduce angular velocity $\bar{\Omega}=\sqrt{{8M_{ADM}}/{3\pi}} \Omega$ (middle panel) and the reduce total horizon area $\bar{A_H}=\sqrt{{27}/(256\pi M_{ADM}^3)}(A_H^{BR}+A_H^{BH})$ (right panel). The arrow indicates the variation of these quantities along the line $j_{BS}^2 = 1$ in Fig.\ref{fig:bsequilibrium}, from $\kappa_2 = 0$ to $\kappa_1 = 1$.}
\label{fig:redspin1}
\end{figure}

The  parameter $\mathcal{A}$ is
\begin{equation}
    \mathcal{A} = 2 \pi L^2 \frac{(\kappa_1-\kappa_2)^2(\kappa_1+\kappa_2-1)}
{\kappa_2(1-\kappa_2)^2}\sqrt{\frac{\kappa_1}{\kappa_2}(\kappa_1 \kappa_2 -(\kappa_1+\kappa_2-1)^2)} \ ,
\label{areabeta}
\end{equation}
and the Gibbs free energy of this system is given by
\begin{equation}
    W=\frac{M^{BH} + M^{BR}}{3} - \frac{\delta}{8 \pi}\mathcal{A}\ ,
\end{equation}
taking into account the constraints (\ref{equalav}) and (\ref{equalt}).

The appropriate set of thermodynamical variables is $({T_H},\Omega,\mathcal{A})$,  where 
\begin{eqnarray}
     {T_H} = \frac{(1-\kappa_2)^2 \kappa_2^3}{2\pi L \kappa_1}\sqrt{\frac{1-\kappa_2}{1-\kappa_1}}\frac{1}{(\kappa_1+\kappa_2-1)
(\kappa_1-\kappa_1^2+\kappa_2-\kappa_2^2)} \\ \nonumber  \times \frac{1}{\sqrt{(2(\kappa_1+\kappa_2-1)
(\kappa_1 \kappa_2 -(\kappa_1+\kappa_2-1)^2))}} \ ,
\end{eqnarray}
is the temperature of the system that results from ${T_H}=T^{BR}_H=T^{BH}_H$;
\begin{eqnarray}
     \Omega = \frac{(1-\kappa_2) \kappa_2}{ L (\kappa_1+\kappa_2-1)}\sqrt{\frac{\kappa_2
(\kappa_1\kappa_2-\kappa_1+\kappa_1^2-\kappa_2+\kappa_2^2)}
{2\kappa_1(\kappa_1+\kappa_2-1)(\kappa_1-\kappa_1^2+\kappa_2-\kappa_2^2)^2}},
\end{eqnarray}
is angular velocity of the black ring and the black hole, $i.e.$ $\Omega=\Omega^{BR}=\Omega^{BH}$, and $\mathcal{A}$ is defined in (\ref{areabeta}). The entropy is
\begin{eqnarray*}
    S &=&  -\frac{\partial{W}}{\partial{{T_H}}}\Bigg|_{\Omega,\mathcal{A}} 
= \pi^2 L^3  \frac{(1 - \kappa_1)^2 \kappa_1}{(1 - \kappa_2)^3 \kappa_2^3}
(\kappa_1+\kappa_2-1)(2\kappa_1+\kappa_2-1)(\kappa_1-\kappa_1^2+\kappa_2-\kappa_2^2)
\\ &&  \qquad \qquad \qquad  \times \sqrt{-\frac{(1 - \kappa_2)(1 - 2 \kappa_1 + \kappa_1^2 - 
  2 \kappa_2 + \kappa_1 \kappa_2 + \kappa_2^2)}{(1 - \kappa_1)(\kappa_1+\kappa_2-1)}} = \frac{1}{4}(A^{BR}_H+A^{BH}_H) \ ;
\end{eqnarray*}
so, once again, the Bekenstein-Hawking area law is recovered. A similar calculation shows that 
$$J = -\frac{\partial{W}}{\partial{\Omega}}\Bigg|_{{T_H},\mathcal{A}} 
= J^{BR}+J^{BH}, \quad \quad {P} = \frac{\partial{W}}{\partial{\mathcal{A}}}\Bigg|_{{T_H},\Omega} 
= -\frac{\delta}{8\pi}.$$

From the above results, one may again conclude that the thermodynamical mass $\mathcal{M}$, computed from $\mathcal{M}=W+T_H S+\Omega J$, yields exactly \eqref{thermomass}
$$\mathcal{M}= M^{BH} + M^{BR}-\frac{\delta}{8\pi}\mathcal{A}=M_{ADM}-E_{int} \ .$$

\begin{figure}
\centering
\includegraphics[width=0.4\textwidth]{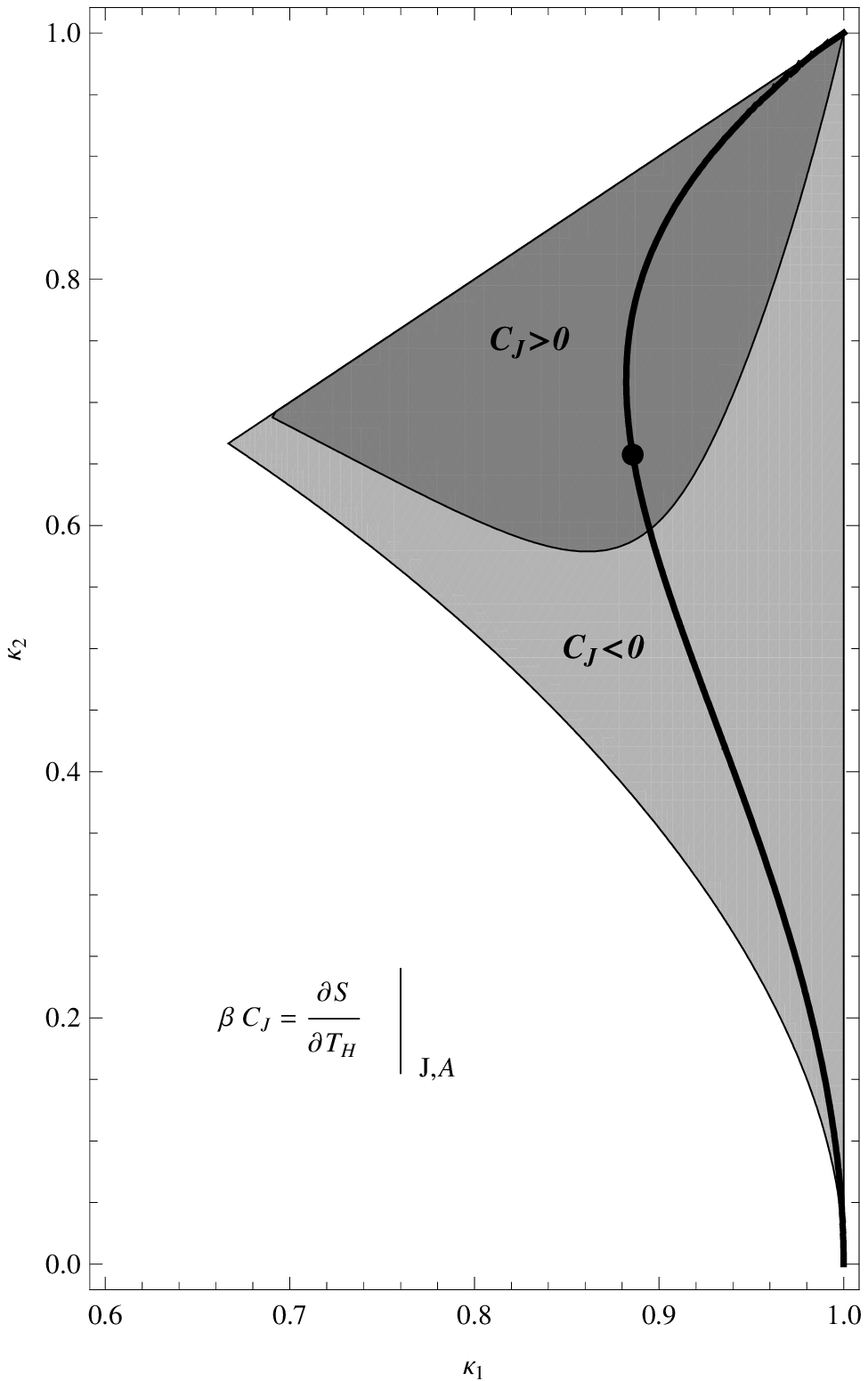} \ \ \ \ \ \ \ \ \ \ 
\includegraphics[width=0.4\textwidth]{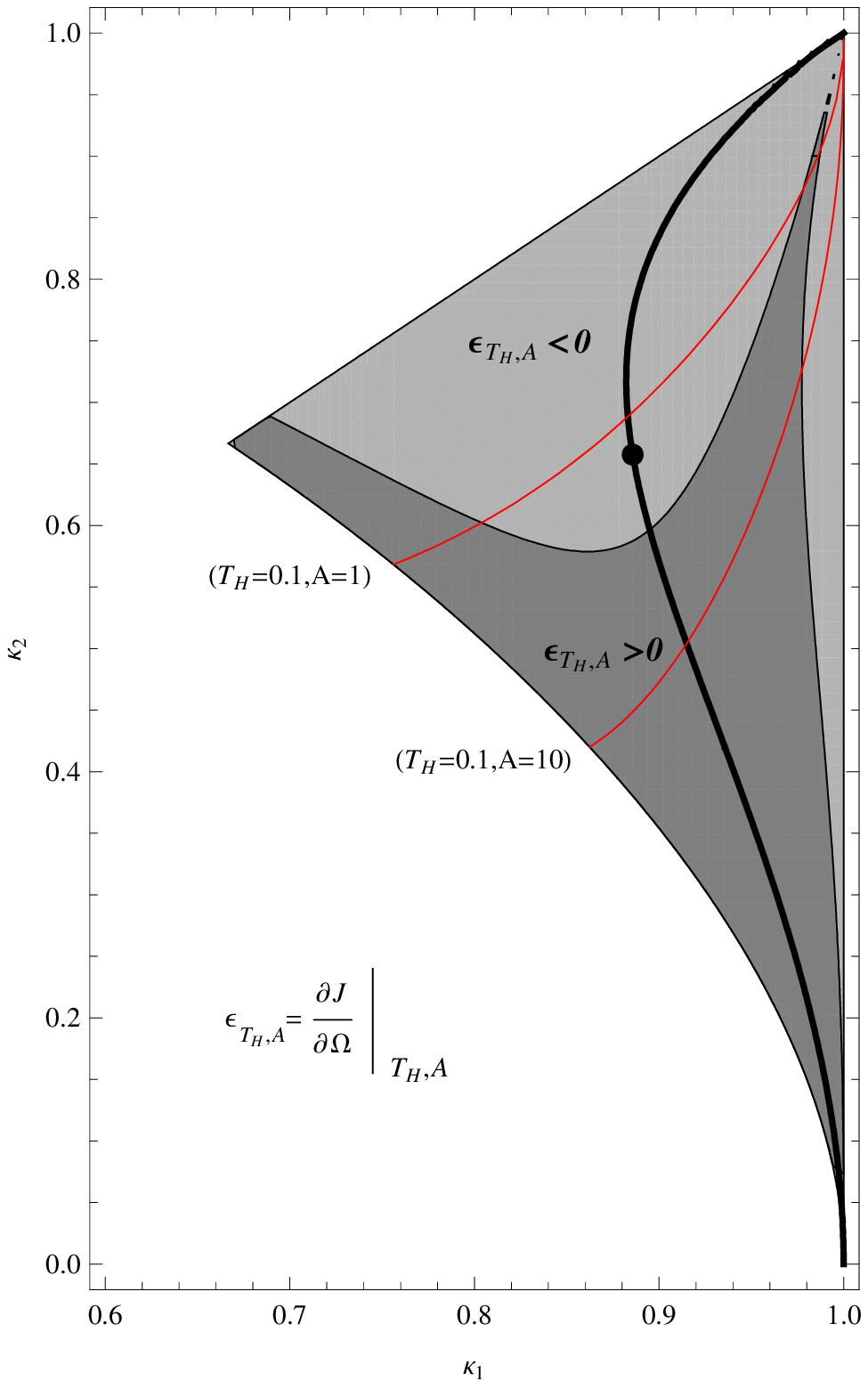}
\caption{Left panel: sign of the specific heat. Right panel: sign of the isothermal moment of inertia.} 
\label{bscjet}
\end{figure}

\begin{figure}
\centering
\includegraphics[width=0.4\textwidth]{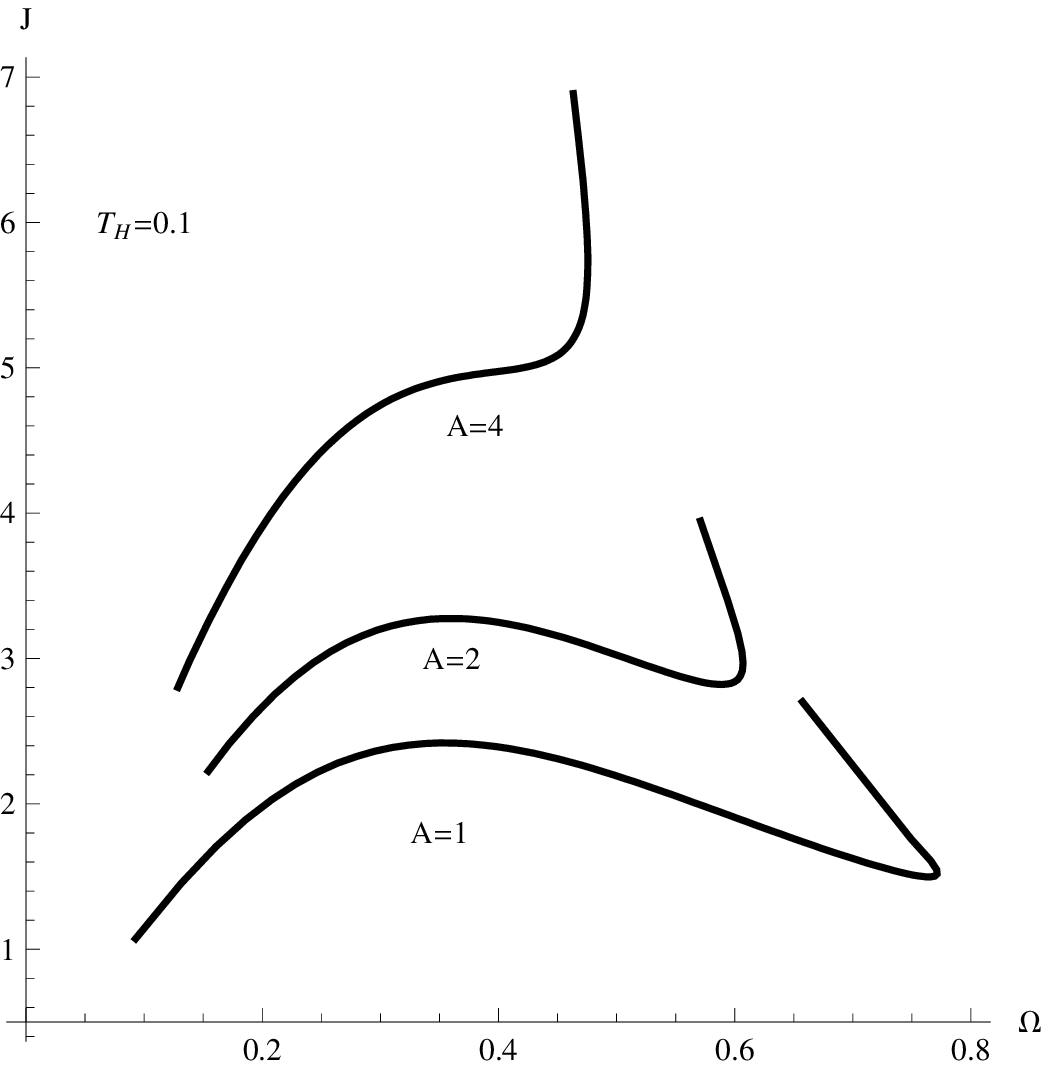} \ \ \ \ \ 
\includegraphics[width=0.4\textwidth]{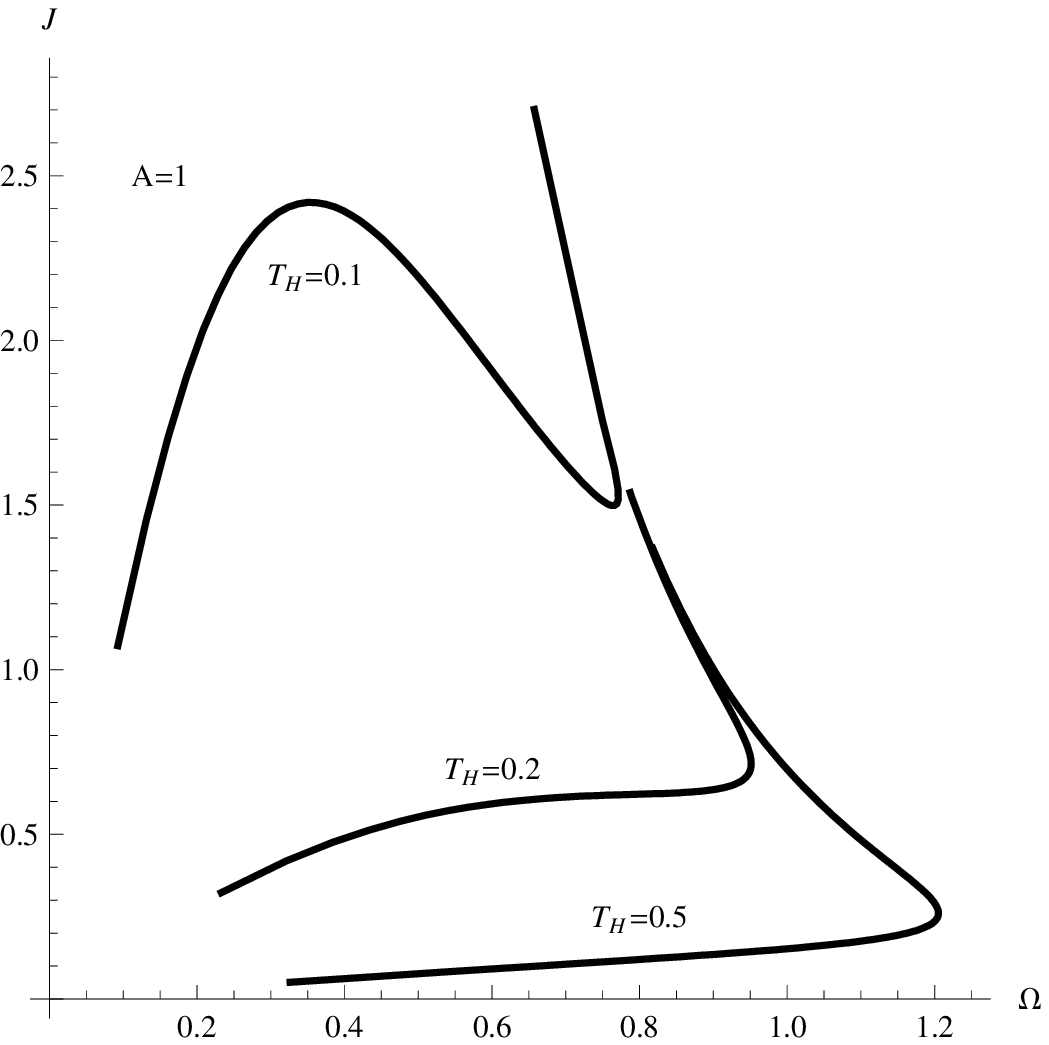}
\caption{Isotherms - at constant $T_H,\mathcal{A}$ - for the $S^1$ rotating black ring. Each $(J,T_H,\mathcal{A})$ generally corresponds to three values of $\Omega$ for low values of $T_H$ and $\mathcal{A}$. For high values of $T_H$ and/or $\mathcal{A}$, $\epsilon_{T_H,\mathcal{A}}$ just changes sign once.}  
\label{fig:IsoBS}
\end{figure}

Finally, the thermodynamical stability of the black Saturn is analysed in Fig. \ref{bscjet}, where the signs of the isothermal moment of inertia and specific heat at constant angular momentum are plotted. Again, as for the single and double Kerr and the $S^1$ rotating ring,  there is no region in parameter space wherein both quantities are positive, and hence no region wherein the solution is thermodynamically stable in the grand canonical ensemble. 

\section{Final Remarks}

In this paper, we have further developed the proposal in \cite{Herdeiro:2009vd} for the thermodynamical description of asymptotically flat solutions with conical singularities. 
The main observation in \cite{Herdeiro:2009vd} is that this description yields 
the Bekenstein-Hawking formula for the entropy which is obtained by differentiating 
the Gibbs free energy. In the Euclidean approach to quantum gravity, the Gibbs 
free energy is obtained from the Euclidean gravitational action. 
Observe that in previous approaches, deviations from the Bekenstein-Hawking 
formula were obtained, in solutions with non-connected event horizons \cite{Costa:2000kf}. 
Here, we have considered various examples of stationary solutions, and showed that the 
description gives the natural results; thus, the thermodynamical angular momentum, obtained again
 by differentiating the Gibbs free energy coincides with the ADM angular momentum, 
 again in contrast with previous descriptions \cite{Costa:2009wj}. Given these results, 
 which support the idea that we have a reliable thermodynamical description of these solutions, 
 we have also considered their thermodynamical stability. We have found that there 
 is no point in the parameter space for which the appropriate Hessian matrix is positive definite. 
 Thus these solutions are always unstable in the grand canonical ensemble. 

 It is worthwhile remarking that, similarly to the static case in \cite{Herdeiro:2009vd},
the  location of the conical singularity was a matter of choice. After a suitable re-scaling, all solutions considered in this work have an alternative interpretation as non-asymptotically flat black objects (i.e the conical singularity may be chosen to extend to spatial infinity). The asymptotic spacetime then corresponds to a cosmic string spacetime for $d=4$, or to its higher dimensional analogue (a membrane for $d=5$). The conical deficit/excess $\delta^\infty$ differs from $\delta$ of the corresponding  solution with the conical singularity having a compact support in the bulk, the relation between these two quantities being
\begin{eqnarray}
\left(1-\frac{\delta}{2\pi}\right)\left(1-\frac{\delta^\infty}{2\pi}\right)=1 \ .
\end{eqnarray}
The formalism proposed in Section \ref{formalism} can easily be generalised to this situation and the  thermodynamical behaviour of the system, in particular its instability, should be independent of the choice for the location of the conical singularities.

\section*{Acknowledgements}

E.R. would like to thank M. J. Rodriguez for useful discussions.
C.H. is supported by a Ci\^encia 2007 research contract.  
The work of E.R. was supported by a fellowship from the Alexander von Humboldt Foundation.  
C. R. is funded by FCT through Grant
No. SFRH/BD/18502/2004. This work has been further supported
by the FCT Grant No. CERN/FP/109306/2009.


\end{document}